\documentclass[a4paper,11pt]{article}

\usepackage{jcappub} 

\usepackage[utf8]{inputenc}

\usepackage[T1]{fontenc} 

\usepackage{subcaption}
\usepackage{bm}
\usepackage{slashed}
\usepackage{comment}
\usepackage{soul}

\usepackage{tikz-feynman}
\usepackage{tikz}
\usetikzlibrary{decorations.pathmorphing}
\usetikzlibrary{decorations.pathmorphing} 

\title{\boldmath Observing Leptogenesis in Action with Gravitational Waves}

\author[a,b]{Hitoshi Murayama}
\author[a,c]{Bea Noether}
\author[a,d]{Jan Schütte-Engel}

\affiliation[a]{Department of Physics, University of California, Berkeley, California 94720, USA}
\affiliation[b]{Kavli Institute for the Physics and Mathematics of the
  Universe (WPI), University of Tokyo,
  Kashiwa 277-8583, Japan}
\affiliation[c]{Ernest Orlando Lawrence Berkeley National Laboratory, Berkeley, California 94720, USA}
\affiliation[d]{RIKEN iTHEMS, Wako, Saitama 351-0198, Japan}

\emailAdd{hitoshi@berkeley.edu}
\emailAdd{bea\_noether@berkeley.edu}
\emailAdd{janschue@berkeley.edu}

\abstract{
Leptogenesis is arguably the best motivated theory of baryogenesis given the discovery of finite neutrino masses, yet its experimental test is elusive given its high energy scale.
We discuss gravitational waves (GWs) produced via graviton bremsstrahlung in right-handed neutrino decays during leptogenesis. 
The presence of right-handed neutrinos in the early universe can lead to a period of early matter domination. In this context, the resultant GW spectrum scales quadratically with the right-handed neutrino mass, while its peak frequency scales inversely with the Yukawa coupling. Detecting such a spectrum would provide strong evidence for leptogenesis and the existence of heavy right-handed neutrinos.
We also discuss how the GW spectrum emitted from the thermal plasma is altered by an era of early matter domination. We show that it can mimic the effects of additional relativistic degrees of freedom and a higher reheating temperature, and that information from the graviton bremsstrahlung GW spectrum can break this degeneracy. 
}

\newcommand{\trh}{t_{\rm rh}}
\newcommand{\tMD}{t_{\rm MD}}
\newcommand{\tM}{t_{\rm M}}
\newcommand{\tD}{t_{\rm D}}
\newcommand{\tZ}{t_{0}}

\newcommand{\Trh}{T_{\rm rh}}
\newcommand{\TMD}{T_{\rm MD}}
\newcommand{\TM}{T_{\rm M}}
\newcommand{\TD}{T_{\rm D}}
\newcommand{\TZ}{T_{0}}

\newcommand{\arh}{a_{\rm rh}}
\newcommand{\aMD}{a_{\rm MD}}
\newcommand{\aM}{a_{\rm M}}
\newcommand{\aD}{a_{\rm D}}
\newcommand{\aZ}{a_{0}}

\newcommand{\Hrh}{H_{\rm rh}}

\newcommand{\gsrh}{g_{*}^{\rm rh}}

\newcommand{\gs}{g_{*}}

\newcommand{\Omgw}{\Omega_{\rm gw}} 

\newcommand{\rhogwZ}{\rho_{\rm gw}^{0}}
\newcommand{\rhogw}{\rho_{\rm gw}}

\newcommand{\rhogammaZ}{\rho^{0}_\gamma}

\newcommand{\rhoNrh}{\rho_{\rm N}^{\rm rh}}
\newcommand{\rhoNMD}{\rho_{\rm N}^{\rm MD}}
\newcommand{\rhoNM}{\rho_{\rm N}^{\rm M}}
\newcommand{\rhoND}{\rho_{\rm N}^{\rm D}}

\newcommand{\rhoN}{\rho_{\rm N}}

\newcommand{\rhoSMrh}{\rho_{\rm SM}^{\rm rh}}
\newcommand{\rhoSMMD}{\rho_{\rm SM}^{\rm MD}}
\newcommand{\rhoSMM}{\rho_{\rm SM}^{\rm M}}

\newcommand{\rhoSM}{\rho_{\rm SM}}

\newcommand{\rhoZ}{\rho^{0}}

\newcommand{\kD}{k_{\rm D}}
\newcommand{\kZ}{k_{0}}

\newcommand{\OmegagammaZ}{\Omega^{0}_\gamma}

\newcommand{\gsrho}{g_{*\rho}}

\newcommand{\gssZ}{g_{*s}^{\rm 0}}
\newcommand{\gss}{g_{*s}}

\newcommand{\gsSM}{g_{*{\rm SM}}}
\newcommand{\gsN}{g_{*{\rm N}}}

\newcommand{\fgw}{f_{\rm gw}}

\newcommand{\Ggw}{G_{\rm gw}}

\newcommand{\nB}{n_{\rm B}}
\newcommand{\nF}{n_{\rm F}}

\newcommand{\SU}{\text{SU}}
\newcommand{\U}{\text{U}}

\begin{document}

\maketitle

\section{Introduction}

The baryon asymmetry of the universe is literally our existential question. It becomes a serious issue when cosmic inflation is adopted as the framework that resolves the flatness, horizon~\cite{Guth:1980zm}, and monopole problems~\cite{Sato:1981qmu}, and provides the mechanism for generating quantum fluctuations that seed structure formation~\cite{Starobinsky:1982ee}. 
Once inflation ends and the universe reheats, any pre-existing baryon asymmetry is wiped out. Therefore, it must be generated after inflation by a microphysical mechanism known as baryogenesis.
It requires (1) baryon number violation, (2) CP violation, and (3) the departure from chemical equilibrium, as pointed out originally by Sakharov~\cite{Sakharov:1967dj}. The Standard Model (SM) of particle physics cannot generate the baryon asymmetry on its own.

Many mechanisms have been proposed for the baryogenesis, see for example Ref.~\cite{Riotto:1998bt} for a review. Arguably the most motivated one today is leptogenesis \cite{Fukugita:1986hr}. We have discovered that neutrinos $\nu_{1,2,3}$, described as massless particles in the SM, actually have non-zero masses at least seven orders of magnitudes smaller than the electron~\cite{Super-Kamiokande:1998kpq}. The best explanation for such tiny masses is that superheavy right-handed Majorana neutrinos $N_{1,2,3}$, new particles beyond the SM, generate the tiny masses of the observed neutrinos~\cite{Minkowski:1977sc,Yanagida:1979as,Gell-Mann:1979vob}. This mechanism is known as the seesaw mechanism because the masses of the observed neutrinos decrease as the masses of the right-handed neutrinos increase. It requires right-handed neutrinos to have Yukawa couplings with the observed neutrinos and the Higgs boson $Y_{i\alpha} \bar{\nu}_i N_\alpha h$. 
If the lightest of the right-handed neutrinos $N_1$ is produced after  reheating, its decay naturally picks up CP violation from the Yukawa couplings, and creates a lepton asymmetry ${\rm BR}(N_1 \rightarrow \nu_i h) \neq {\rm BR}(N_1 \rightarrow \bar{\nu}_i h)$. Mildly small Yukawa couplings $Y \lesssim 0.2\,\left(M_1/(10^{15}\,{\rm GeV})\right)^{1/2}$, where $M_1$ is the mass of $N_1$, make the right-handed neutrino decay out of equilibrium.  The lepton asymmetry gets subsequently partially converted into a baryon asymmetry by the anomaly in the electroweak interaction~\cite{Kuzmin:1985mm}.

Testing leptogenesis, however, is notoriously difficult. This is simply because the right-handed neutrino needs to be very heavy $10^8~{\rm GeV} \lesssim M_1 \lesssim 10^{15}~{\rm GeV}$ \cite{Buchmuller:2004nz}, which is beyond the reach of any imaginable accelerator. Given that a direct test is not possible, we need to look for circumstantial evidence. This includes searches for lepton-number violation in neutrinoless double beta decay  $0\nu\beta\beta$ and CP violation in neutrino oscillations. However, since such signals are circumstantial, additional evidence is required.

Contrary to laboratory searches for physics at higher energies, which rely on higher-dimensional operators and are hence suppressed by their energy scales, GWs are more sensitive to physics the higher the energy scales are. One of the authors (HM), with collaborators \cite{Dror:2019syi}, pointed out that the masses of right-handed neutrinos likely originate from the Higgs mechanism of new gauge groups such as $\U(1)_{B-L}$. If the symmetry breaking occurs at an energy scale $v_{B-L} < \Trh$, where $\Trh$ is the reheating temperature and $v_{B-L}$ the $\U(1)_{B-L}$ symmetry breaking scale, a network of cosmic strings can form via the Kibble–Zurek mechanism \cite{Zurek:1985qw,Murayama:2009nj}, analogous to the Abrikosov flux of magnetic fields in Type-II superconductors. The cosmic string network then keeps simplifying itself emitting GWs in a nearly scale-invariant fashion. The resulting signal of GWs is within the range of sensitivity of future space missions~\cite{Dror:2019syi}. It is  nearly scale-invariant and is proportional to $v_{B-L}^2 / m_{\rm Pl}^2$, where $m_{\rm Pl}$ is the Planck mass $m_{\rm Pl}=1.22\times 10^{19}\,$GeV.  Such a cosmic string network is inevitable for $\SU(3)_C \times \SU(2)_L \times \U(1)_Y \times \U(1)_{B-L}$, and also possible for all other gauge groups $\SU(3)_C \times \SU(2)_L \times \SU(2)_R\times \U(1)_{B-L}$, $\SU(4)_{PS} \times \SU(2)_L \times \U(1)_Y$, or $\SU(5)\times \U(1)_X$ which are free from magnetic monopoles with the minimum rank. 

Yet, it would be ideal if we could observe the decay of right-handed neutrinos as it is the process that creates the asymmetry. In this paper, we discuss whether gravitons emitted in a bremsstrahlung-like process during the decay of right-handed neutrinos can be detected. 
Such a detection could be regarded as direct observation of the decay of heavy right-handed neutrinos, although it would not constitute a proof that the process generated a lepton asymmetry. 
However, when combined with other circumstantial evidence, such an observation would make the case for leptogenesis far more convincing. In fact, the emitted gravitons eventually form a stochastic, incoherent background with a characteristic frequency spectrum serving as a smoking-gun signature of right-handed neutrino decays. 
The magnitude of the gravitational wave spectrum is proportional to $M_1^2$ and the spectrum peaks at high frequencies, with the peak frequency scaling inversely with the Yukawa coupling. It is encouraging that serious efforts are underway to develop technologies for gravitational wave detection in this high-frequency regime.
The presence of heavy right-handed neutrinos in the early universe can induce an era of early matter domination which will also inevitably modify the GW spectrum from the thermal plasma, which is know as Cosmic Gravitational Microwave Background (CGMB)~\cite{Ghiglieri:2015nfa,Ghiglieri:2020mhm,Ringwald:2020ist}. Together with the bremsstrahlung GW spectrum, it would provide additional evidence that a heavy particle once dominated the universe.

The GW spectrum from graviton bremsstrahlung in right-handed neutrino decays during leptogenesis has also been calculated in Refs.~\cite{Datta:2024tne,Choi:2025hqt}. Our results differ in two ways from their results. First, we do not agree with their graviton bremsstrahlung decay rate calculation. We give more details on the disagreement in appendix~\ref{sec:Matrix_element}. Second, the authors did not include a phase of early matter domination which can significantly alter the GW spectrum. 
Note that the impact of an era of early matter domination caused by heavy right-handed neutrinos on the inflationary GW spectrum has been explored in Refs.~\cite{Borboruah:2025hai,Chianese:2025mll}, but not in the context of graviton bremsstrahlung from right-handed neutrino decays and the CGMB.

Our paper is organized as follows. In Sec.~\ref{sec:leptogenesis} we review the process of leptogenesis and discuss the connection to right-handed neutrinos and the seesaw mechanism. Sec.~\ref{sec:cosmological_model} covers the cosmological model with heavy right-handed neutrinos. The main results of this paper are presented in Sec.~\ref{sec:GW_spectrum}. In Sec.~\ref{Sec:GW_decay} we calculate the GW spectrum from graviton bremsstrahlung in right-handed neutrino decays during leptogenesis. Then in Sec.~\ref{Sec:GW_thermal_plasma} we calculate the CGMB spectrum with a phase of early matter domination. We conclude in Sec.~\ref{sec:conclusion}.

In this paper we use the metric convention $\eta_{\mu\nu}=(+,-,-,-)$ and we use natural units, i.e., $\hbar=c=k_{\rm B}=1$.

\section{Right-handed neutrinos and leptogenesis}\label{sec:leptogenesis}

The existence of right-handed neutrinos is well-motivated for two reasons. (1) They provide a natural mechanism for generating small active neutrino masses without requiring unnaturally small couplings (seesaw mechanism), and (2) their CP violating decays can produce a lepton asymmetry that is subsequently converted to a baryon asymmetry by the electroweak sphaleron process (leptogenesis). They are also motivated once the SM gauge groups are embedded into $SO(10)$ or $E_6$ grand unified theories. Even without a full-fledged grand unification, an additional gauge group such as $U(1)_{B-L}$ or $SU(2)_R$ necessitates the existence of three right-handed neutrino states to ensure anomaly cancellations.

The Lagrangian of right-handed neutrinos $N_\alpha$ is the following,
\begin{align}
	{\cal L} &= {\cal L}_{\rm SM} + \bar{N}_\alpha i {\slash\!\!\!\partial }N_\alpha - \left[ \frac{1}{2} M_\alpha N_\alpha N_\alpha + y_{\alpha i} N_\alpha L_i H + c.c. \right].
\end{align}
Here, we used the two-component notation for the fermion fields, and the basis where the right-handed neutrino mass matrix is real, positive, and diagonal without loss of generality. 

\begin{figure}
	\centering
	\includegraphics[width=0.6\textwidth]{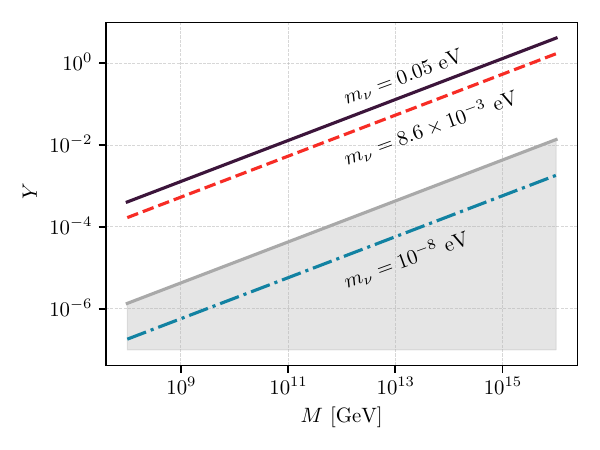}
	\caption{A schematic plot for a normal ordered active neutrino mass spectrum. We use the single generation seesaw equation $m_\nu=(Yv)^2/M$ in order to depict what parameters yield which active neutrino masses. The black solid line corresponds to an active neutrino mass of $m_\nu=0.05\,$eV, the red dashed line to $m_\nu=8.6\times10^{-3}\,$eV and the blue dash-dotted line to $m_\nu=10^{-8}\,$ eV. The gray shaded region indicates the parameter space that leads to an era of early matter domination. }
	\label{fig:seesaw_parameters}
\end{figure}

By integrating out the right-handed neutrinos, we obtain an effective dimension-five operator using the matrix $y$,
\begin{align}
	{\cal L}_5 &= \frac{1}{2} \sum_\alpha y_{\alpha i} y_{\alpha j} \frac{1}{M_\alpha} (L_i H) (L_j H),
\end{align}
or
\begin{align}
	(m_\nu)_{ij} = \sum_\alpha y_{\alpha i} \frac{1}{M_\alpha} y_{\alpha j} \langle H \rangle^2
	= (y^T M^{-1}\,y)_{ij} v^2,
\end{align}
where $v = \langle H \rangle = 174$~GeV.
Applying this to a single flavor, it yields an expression for the active neutrino mass
\begin{align}
	m_\nu 
    = \frac{(Y v)^2}{M} = 0.03~{\rm eV}\ Y^2 \frac{10^{15}~{\rm GeV}}{M},
    \label{eq:seesaw_single_flavor}
\end{align}
Here, we are ignoring the renormalization-group effects which enhance the neutrino mass by gauge interactions while suppressing it by Yukawa interactions. We are only interested in orders of magnitude here. In Fig.~\ref{fig:seesaw_parameters} we show a neutrino mass spectrum in the $M$ and $Y$ parameter-space based on the single flavor Eq.~\eqref{eq:seesaw_single_flavor}.

We assume in this paper that all $N_\alpha$ have similar masses that can be as large as $10^{15}$~GeV. $N_{2,3}$ have Yukawa couplings of $O(1)$, while $N_1$ has smaller Yukawa couplings $y_{1i} \lesssim O(10^{-2})$ to avoid the washout of the lepton asymmetry as we will explain later. This is not a problem for the seesaw mechanism, as the lightest active neutrino may well be (nearly) massless, i.e., much lighter than the other two neutrinos. Note that the recent DESI data  \cite{DESI:2024mwx} suggests that the sum of the three neutrino masses is less than $\sum m_\nu < 0.072~$eV (95\% CL) and prefers the normal mass ordering. Therefore, a possible mass spectrum is
\begin{align}
	m_{\nu_1} \ll m_{\nu_2} = \sqrt{\Delta m^2_{12}} \approx 8.6 \times 10^{-3}~{\rm eV} < m_{\nu_3} = \sqrt{\Delta m^2_{12}+\Delta m^2_{23}} \approx 0.05~{\rm eV}.
\end{align} 
The scenario of interest is that the universe is reheated to $\Trh > M$ and the gauge interactions such as $U(1)_{B-L}$ or $SU(2)_R$ quickly equilibrate the thermal plasma, including the right-handed neutrinos. The upper limit on the tensor-so-scalar ratio $r < 0.032$ \cite{Tristram:2021tvh} gives us an upper bound on the inflationary energy density (see, {\it e.g.}\/, \cite{Domcke:2015iaa})
\begin{align}
	V_* = \frac{3}{128}\, A_s\, r\, m_{\rm Pl}^4 = (3.3 \times 10^{16}~{\rm GeV})^4\, r < (1.4 \times 10^{16}~{\rm GeV})^4.
\end{align}
The highest possible reheating temperature is achieved through instantaneous reheating, which converts the inflationary energy density entirely into thermal energy
\begin{align}
	\frac{\pi^2}{30}\,g_* \, T_{\rm rh,max}^4 = V_*, \qquad T_{\rm rh, max} = 5.7 \times 10^{15}~{\rm GeV}.
\end{align}
Here, we assumed $g_* = 112$ as in the SM with three right-handed neutrinos. 

Once thermalized, we assume that $N_1$ is relatively long-lived and decays later at a much lower temperature, generating the lepton asymmetry. The decay rate of the right-handed neutrinos is given by
\begin{align}
	\Gamma_\alpha &= \frac{1}{8\pi} (yy^\dagger)_{\alpha\alpha} M_\alpha \qquad (\mbox{no sum over } \alpha).
\end{align}
Therefore, the temperature at the decay $\TD$ is estimated by,
\begin{align}
	H(\TD) = \left( \frac{8\pi G}{3} \frac{\pi^2}{30} \gsSM \TD^4 \right)^{1/2}  \simeq \frac{2}{3} \Gamma_1,
\end{align}
where $G=\,1/m_{\rm Pl}^2$ is Newton's constant. The decay temperature is then:
\begin{align}
	\TD \simeq 4.3 \times 10^{11}~{\rm GeV} \left( \frac{M_1}{10^{15}~{\rm GeV}} \right)^{1/2} \left(\frac{(yy^\dagger)_{11}}{(10^{-4})^2}\right)^{1/2}. 
    \label{eq:T_D}
\end{align}
The CP asymmetry in decays of right-handed neutrinos was computed in full generality in Ref.~\cite{Covi:1996wh},
\begin{align}
	\epsilon_1 &= \frac{\Gamma(N_1 \rightarrow l h) - \Gamma(N_1 \rightarrow l^* h^*)}{\Gamma(N_1 \rightarrow l h) + \Gamma(N_1 \rightarrow l^* h^*)}
	= \frac{1}{8\pi} \sum_{\beta \neq \alpha} f(x_\beta) \frac{\Im m[(yy^\dagger)_{\beta,1}]^2}{(yy^\dagger)_{11}},
\end{align}
where
\begin{align}
	f(x) = \sqrt{x} \left[ 1- (1+x) \ln \frac{1+x}{x} \right], \qquad x_\beta = \frac{M_\beta^2}{M_1^2}\ .
\end{align}
Assuming all three masses are comparable, $|f(x)| \sim 0.1$--0.5, and $\epsilon_1 \simeq \Im m[(y_{\beta j})^2]/8\pi$ (in the basis where $y_{1i}$ are all real), with $\beta=2$ or 3 and $j=1,2,3$, which can be as large as 0.01. 

We assume that the lepton asymmetries generated by $N_{2,3}$ decays are washed out, and the gravitons emitted as bremsstrahlung during their decays are redshifted away. The final lepton asymmetry and gravitons are those generated by the $N_1$ decays. In the following we will therefore focus on one right-handed neutrino generation only. 
It would be interesting to combine our study on GWs with a more detailed analysis on leptogenesis, though this lies beyond the scope of the present work. 

The generated lepton asymmetry from $N_1$ can be washed out due to scattering processes in the plasma, $\sigma( l H \rightarrow l^* H^*) \simeq 1/(8\pi M_1^2)$. In order to avoid washout, the scattering rate $\Gamma \simeq \sigma T_D^3$ must be below the expansion rate at the time of the decay, $H \simeq Y^2 M_1 / 8\pi$ where we define $Y\equiv\sqrt{(yy^\dagger)_{11}}$. Therefore, we need $Y \lesssim ( (8\pi)^3 M_1^2 / m_{\rm Pl}^2)^{3/4} \simeq 0.03$. We conservatively require $Y < 10^{-2}$ in our study and use $Y \simeq 10^{-4}$ for numerical estimates. Then we find
\begin{align}
    Y_L \approx \epsilon \frac{T_D}{M_1}
    \approx 5 \times 10^{-6} \Im m(y_{\beta j})^2 \frac{|f(x)|}{0.3} \left( \frac{10^{15}~{\rm GeV}}{M_1}\right)^{1/2} \left(\frac{Y}{10^{-4}}\right),
\end{align}
which makes it clear that we can match the observed value $Y_L=\frac{79}{28}Y_B=2.42\times 10^{-10}$.

The generated asymmetry can in principle also be washed out by inverse decays. However, here we assume that the time of the decay $\tD>\tM$, where $\tM$ is the time when the right-handed neutrinos become non-relativistic. In that scenario inverse decays do not play a role because they are kinematically not possible.
Clearly some level of washout could be tolerated and there are also flavor-dependencies to protect the overall asymmetry.
The framework we use here is simply a benchmark where leptogenesis is guaranteed to work. We will focus our study on the GW signatures within this framework.

Note that the Davidson--Ibarra bound \cite{Davidson:2002qv} limits the right-handed neutrino mass from below $M_1 \gtrsim 10^9$~GeV to obtain a sufficient CP violation. We are interested in a much higher mass and hence there is no problem to create a sufficient asymmetry. From now on we will refer to $M_1$ as just $M$ for convenience.

\section{Cosmological model}\label{sec:cosmological_model}

In this section, we outline the cosmological model used to describe the decay of right-handed neutrinos. We assume that, after reheating (denoted 'rh'), the thermal plasma in the universe was in thermal equilibrium. Furthermore, all SM particles, as well as the right-handed neutrinos in the plasma, are assumed to be relativistic.

In order to have a plasma in thermal equilibrium we need $\Gamma_g>\Hrh$, where the rate $\Gamma_g$ corresponds to the interaction rate of the particles in the plasma which keeps the particles in thermal equilibrium. Parametrically this translates to $g>\sqrt{\Trh/m_{\rm Pl}}$, where $g$ is the coupling. For all SM particles $g$ is the gauge couplings of the SM which are on the order of $\mathcal{O}\left(10^{-1}\right)$ at high temperatures. It is therefore clear that the SM particles can be in thermal equilibrium even at ultra high energies. We can also assume that the right-handed neutrinos are in thermal equilibrium. The Yukawa couplings that we consider are to small in order to achieve that. However, the right-handed neutrinos can remain in thermal equilibrium through gauge interactions.

\begin{figure}
	\centering
	\includegraphics[width=0.9\textwidth]{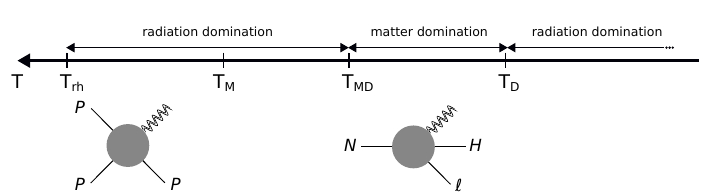}
	\caption{Sketch of the cosmological history. After reheating the universe is radiation dominated. Thermal fluctuations in the plasma produce gravitons through $2\to 2$ scattering processes, giving rise to the CGMB. Particles from the thermal plasma are depicted with '$P$' in the diagram. At $\TM=M$ the right-handed neutrinos become non-relativistic and from $\TMD$ until $\TD$ the universe is matter dominated. After $\TD$ the universe is radiation dominated again because most right-handed neutrinos have decayed away. Gravitons are emitted as bremsstrahlung during the right-handed neutrino decays. }
	\label{Fig:cosmological_history}
\end{figure}

The right-handed neutrinos become non-relativistic when the temperature of the plasma drops below their mass, $M$. If the Yukawa coupling of the right-handed neutrino is not too large, the universe can become matter dominated after some time. We denote the onset of this early matter-dominated era as `MD'. Once the right-handed neutrino abundance is depleted by decays, the universe becomes radiation dominated again. We denote this moment with `D'. 
A sketch of the cosmological history is provided in Fig.~\ref{Fig:cosmological_history}.

In the following we describe for each epoch how the energy density, temperature, scale factor and time evolve. 
The energy densities of the SM particles and right-handed neutrinos after reheating are $\rhoSMrh=\pi^2/30\,\gsSM \,\Trh^4$ and $\rhoNrh=\pi^2/30\,\gsN \,\Trh^4$ with $\gsSM=106.75$ and $\gsN=\frac{7}{8}\times 2$.\footnote{We only consider one right-handed neutrino species in order to disentangle the effects caused by multiple species which might decay fast after being produced.} We also define the effective degrees of freedom $\gs$ as a function of temperature to be $\gs=\gsSM+\gsN$ for $T>M$ and $\gs=\gsSM$ for $T\leq M$. Of course, $\gsSM$ decreases at some point, but as long as the temperature remains above the electroweak scale, it is a good approximation to treat it as constant. In this regime, the effective degrees of freedom for the energy and entropy densities are also approximately equal, $\gss \simeq \gsrho \simeq \gs$. 
Before the neutrinos become non-relativistic the energy densities evolve as radiation $\rhoSM,\rhoN\sim a^{-4}$.
From $\TM=M$ onward the right-handed neutrinos are non relativistic while all SM particles are still relativistic. The scale factor at this moment is $\aM=\arh\,\Trh/M\,\left(\gsrh/\gsSM\right)^{1/3}\simeq \arh\,\Trh/M$, where $\arh$ is the scale-factor at reheating.

The evolution equations for SM and right-handed neutrino energy densities are
\begin{eqnarray}
	(\partial_t+3H)\rhoN&=&-\Gamma \rhoN,\label{eq:rho_evolution_MD_N}\\
	(\partial_t+4H)\rhoSM&=&\Gamma \rhoN.\label{eq:rho_evolution_MD_R}
\end{eqnarray}
The solution for $\rhoN$ is,
\begin{eqnarray}
	\rhoN=\rhoNM \left(\frac{\aM}{a}\right)^3\, e^{-\Gamma(t-\tM)},
\end{eqnarray}
while the solution for $\rhoSM$ is
\begin{eqnarray}
	 \rhoSM(t)  = \left(\frac{\aM}{a(t)}\right)^4\rhoSMM+ \Gamma\,\int_{\tM}^{t} dt'\, \rhoN(t')\, \left(\frac{a(t')}{a(t)}\right)^4,
\end{eqnarray}
where we start the integral at $\tM$ because we assume that the decays happen mainly after $\tM$, i.e., $\Gamma \,\tM<1$. In order to get a phase of early matter domination we must also require that the time when the matter and radiation energy densities are equal, $\tMD$, is earlier than $\tD$, the time when the universe becomes radiation dominated again. Writing this constraint out yields
\begin{eqnarray}
    Y^2<0.23\,\frac{M}{m_{\rm Pl}},
    \label{eq:condition_for_EMD}
\end{eqnarray}
where we have used that $\tMD=\tM \left(\aMD/\aD\right)^2$ with 
\begin{eqnarray}
    \tM=\frac{m_{\rm Pl}}{M^2}\frac{1}{2}\left(\frac{8\pi }{3}\frac{\pi^2}{30}\gsrh\right)^{-1/2}.    
\end{eqnarray}
We show the parameter regime where \eqref{eq:condition_for_EMD} is satisfied in gray in Fig.~\ref{fig:seesaw_parameters}.
In the regime $\tM<t<\tMD$ the leading contribution to the SM energy density is $\rhoSM(t)\simeq \left(\aM/a\right)^4\rhoSMM $, while the neutrino energy density scales as $\rhoSM(t)\simeq\left(\aM/a\right)^3\rhoNM $. This leads to $\aMD=\aM\,\gsSM/\gsN$ and $\TMD= \TM \,\aM/\aMD$.

The evolution equation for $\rhoSM$ from the start of early matter domination (MD) to the end (D) can be obtained by integration of Eq.~\eqref{eq:rho_evolution_MD_R},
\begin{eqnarray}
	\rhoSM(t)  
	&=&\left(\frac{\aMD}{a}\right)^4\rhoSMMD+\left(\frac{\aMD}{a}\right)^4 \rhoNMD  \frac{e^{\Gamma \tMD} \left(\Gamma \left(\frac{5}{3},\Gamma \tMD\right)-\Gamma \left(\frac{5}{3},\Gamma t\right)\right)}{ (\Gamma\tMD)^{2/3}},
    \label{eq:rhoSM_MD_t0_D}
\end{eqnarray}
where $\Gamma(\cdot,\cdot)$ is the incomplete gamma function.

\begin{figure}
	\centering
	\includegraphics[width=0.6\linewidth]{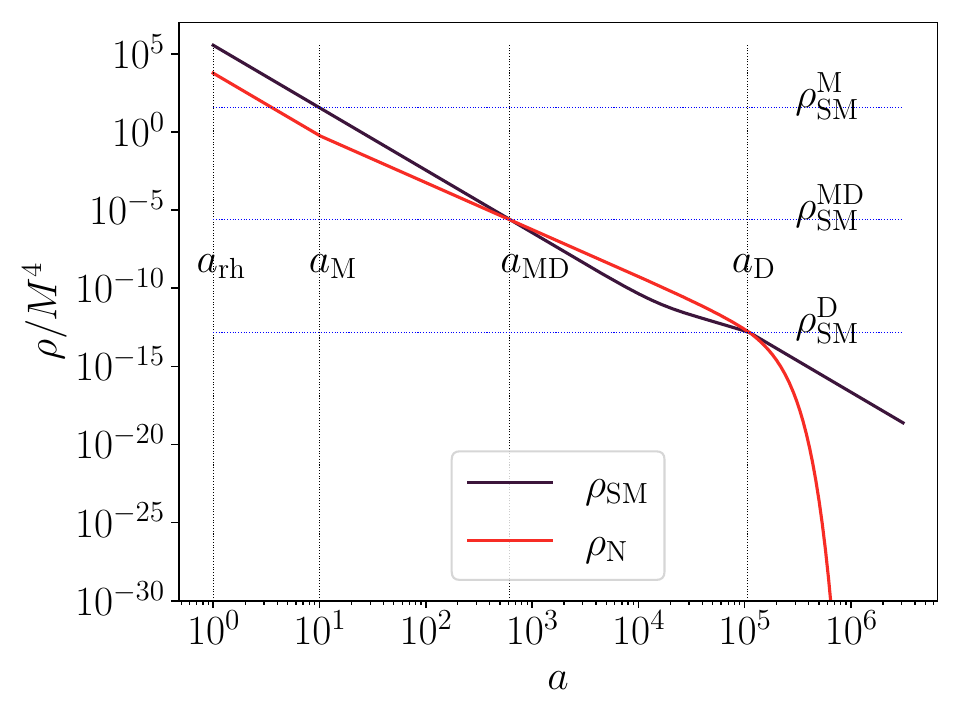}
	\caption{Evolution of the energy densities for $\Trh=10^{16}$ GeV, $M=10^{15}$ GeV and $Y=10^{-4}$. In black we show the evolution of the SM energy density and in red the evolution of the right-handed neutrino energy density. The right handed neutrinos become non-relativistic at $\aM$. Early matter domination starts at $\aMD$ and ends at $\aD$.}
	\label{Fig:Evolution_energy_density_Temperature}
\end{figure}

Early matter domination stops when the right handed neutrino and SM energy densities are equal again. This happens when $t=\tD=\Gamma^{-1}$ what implies that the scale factor at the end of matter domination is $\aD=\aMD\left(\tD/\tMD\right)^{2/3}$.
The temperature $\TD$ is again defined through the energy density of the relativistic particles at $\tD$.
We can re-write the expression from Eq.~\eqref{eq:T_D},
\begin{eqnarray}
	\TD\simeq M\,\left(\frac{\aMD}{\aD}\right)^{3/4} \left(\frac{\gsN}{\gsSM}\right).
	\label{Eq:TD}
\end{eqnarray}
We show the evolution of the energy densities in Fig.~\ref{Fig:Evolution_energy_density_Temperature}.
In order to produce the plots we have used the full expressions with the Gamma functions which also confirms that the estimates for scale factors and energy densities whose values we indicate with thinner and dotted lines are accurate.
Fig.~\ref{Fig:Evolution_energy_density_Temperature} also confirms our intuition that before the right-handed neutrinos become non-relativistic the temperature scales as $T\sim a^{-1}$. Between $\tMD$ and $\tD$ there is a short period where $T\sim a^{-3/8}$. When the universe becomes radiation dominated again, we get the usual scaling $T\sim a^{-1}$ back.

\section{Gravitational wave spectrum}\label{sec:GW_spectrum}

In the cosmological scenario that we show in Fig.~\ref{Fig:cosmological_history}, GWs will be generated in two different ways. 
Firstly, when the right-handed neutrinos become non relativistic and start to decay, gravitons are generated by from graviton bremsstrahlung. We calculate the corresponding graviton bremsstrahlung from right-handed neutrino decays in Sec.~\ref{Sec:GW_decay}.
Secondly, GWs are produced from thermal fluctuations in the plasma. The contribution is known as the Cosmic Gravitational Microwave background (CGMB) and its dominant contribution is produced at high temperatures. An era of early matter domination will shift the peak of the CGMB spectrum to lower frequencies. We calculate the CGMB spectrum with an early matter dominated phase in Sec.~\ref{Sec:GW_thermal_plasma}.

The evolution equation of the graviton energy density, $\rhogw$, is given by
\begin{align}
	(\partial_t+4H(t))\rhogw(t)=\int \frac{d^3k}{(2\pi)^3}\, 2k \Ggw(k,t),
	\label{eq:rhogw_equation}
\end{align}
where $\Ggw$ is the so called gain term which depends on the processes that generate gravitons. In principle there is also a loss term in Eq.~\eqref{eq:rhogw_equation}. However, such a term will always be much smaller than the gain term since the graviton distribution function is always much smaller than one.

\subsection{Gravitational waves from right-handed neutrino decays}\label{Sec:GW_decay}

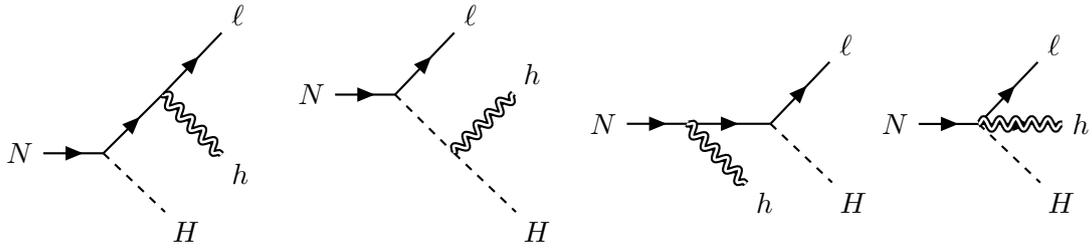
\begin{figure}[t]
	\centering
	\begin{subfigure}[c]{.24\textwidth}
		\begin{tikzpicture}
			\begin{feynman}
				\vertex (a) {\(N\)};
				\vertex [right=1.1cm of a] (b) ;
				\vertex [above right=1.1cm of b] (c);
				\vertex [above right=1.1cm of c] (d) {\(\ell\)};
				\vertex [below right=1.1cm of c] (e) {\(h \)};
				\vertex [below right=1.1cm of b] (f) {\(H \)};
				\diagram* {
					(a) -- [fermion,thick] (b), 
					(b) -- [fermion,thick] (c),
					(c) -- [scalar,thick] (e),
					(c) -- [thick, double distance=1pt,decoration={snake, amplitude=2pt, segment length=6pt}, decorate] (e),
					(c) -- [fermion,thick] (d),
					(b) -- [scalar,thick] (f)
				};
				~~~~~~\end{feynman}
		\end{tikzpicture}
	\end{subfigure}	
	\begin{subfigure}[c]{.24\textwidth}
		\begin{tikzpicture}
			\begin{feynman}
				\vertex (a) {\(N\)};
				\vertex [right=1.1cm of a] (b) ;
				\vertex [above right=1.1cm of b] (c) {\(\ell\)};
				\vertex [below right=1.1cm of b] (d) ;
				\vertex [below right=1.1cm of d] (f) {\(H \)};
				\vertex [above right=1.1cm of d] (e) {\(h \)};
				\diagram* {
					(a) -- [fermion,thick] (b), 
					(b) -- [fermion,thick] (c),
					(b) -- [scalar,thick] (d),
					(d) -- [scalar,thick] (e),
					(d) -- [thick, double distance=1pt,decoration={snake, amplitude=2pt, segment length=6pt}, decorate] (e),
					(d) -- [scalar,thick] (f)
				};
				~~~~~~\end{feynman}
		\end{tikzpicture}
	\end{subfigure}
	\begin{subfigure}[c]{.24\textwidth}
	\begin{tikzpicture}
		\begin{feynman}
			\vertex (a) {\(N\)};
			\vertex [right=1.1cm of a] (b) ;
			\vertex [right=1.1cm of b] (d) ;
			\vertex [below right=1.1cm of b] (c) {\(h\)};
			\vertex [above right=1.1cm of d] (e) {\(\ell\)};
			\vertex [below right=1.1cm of d] (f) {\(H\)};
			\diagram* {
				(a) -- [fermion,thick] (b), 
				(b) -- [fermion,thick] (d),
				(b) -- [scalar,thick] (c),
				(b) -- [thick, double distance=1pt,decoration={snake, amplitude=2pt, segment length=6pt}, decorate] (c),
				(d) -- [fermion,thick] (e),
				(d) -- [scalar,thick] (f)				};
			~~~~~~\end{feynman}
	\end{tikzpicture}
\end{subfigure}
	\begin{subfigure}[c]{.24\textwidth}
		\begin{tikzpicture}
			\begin{feynman}
				\vertex (a) {\(N\)};
				\vertex [right=1.1cm of a] (b) ;
				\vertex [above right=1.1cm of b] (c){\(\ell \)};
				\vertex [right=1.1cm of b] (d) {\(h \)};
				\vertex [below right=1.1cm of b] (e) {\(H \)};
				\diagram* {
					(a) -- [fermion,thick] (b), 
					(b) -- [fermion,thick] (c),
					(b) -- [scalar,thick] (e),
					(b) -- [fermion,thick] (d),
					(b) -- [thick, double distance=1pt,decoration={snake, amplitude=2pt, segment length=6pt}, decorate] (d)
				};
				~~~~~~\end{feynman}
		\end{tikzpicture}
	\end{subfigure}	
	\caption{Feynman diagrams for graviton bremsstrahlung production from right-handed neutrino decay. }
	\label{fig:gw_production_neutrino_decay_main_text}
\end{figure}

In this section we calculate the GW spectrum from graviton bremsstrahlung processes in right-handed neutrino decays. We show the corresponding Feynman diagrams in Fig.~\ref{fig:gw_production_neutrino_decay_main_text}. The gain term is given by:
\begin{eqnarray}
	\Ggw(k,t)=\frac{1}{4k}\int\frac{d^3p}{(2\pi)^3\,2E_p}\frac{d^3q}{(2\pi)^3\,2E_q}\frac{d^3r}{(2\pi)^3\,2E_r} \left|\mathcal{M}\right|^2 f_{\rm N}(p)\,\times
    \nonumber\\
    \times\left(1+\nB(r)\right)\,\left(1-\nF(q)\right)\,
    (2\pi)^4\delta(P-Q-K-R).
	\label{eq:collision_term}
\end{eqnarray}
We use capital letters for the four momenta of the particles. The four momenta of the right-handed neutrino, lepton, scalar and graviton are $P$, $Q$ , $R$ and $K$ respectively. We refer to the magnitude of the three momentum vectors as $|\bm{p}|=p,\,|\bm{k}|=k,\,|\bm{r}|=r,\,|\bm{q}|=q$. The Bose-Einstein and Fermi-Dirac distributions are defined as $n_{\rm B,F}(q)=1/(e^{q/T}\mp 1)$ and the distribution function of the right-handed neutrinos is $f_{\rm N}$.
Note that in principle we should have also added a term $(1+f_{\rm gw})$ in Eq.~\eqref{eq:collision_term}, where $f_{\rm gw}$ is the graviton distribution function. However, $f_{\rm gw}\ll 1$ and hence we can neglect such a term. The distribution functions can also depend on time as the temperature evolves with time but for better readability we suppress the time argument. The matrix element squared that appears in Eq.~\eqref{eq:collision_term} is the spin and helicity averaged matrix element that we derive in appendix~\ref{sec:Matrix_element}.\footnote{Note that the $\left|\mathcal{M}\right|^2$ that appears in Eq.~\eqref{eq:collision_term} is four times the expression from  Eq.~\eqref{eq:Msq_averaged} because there we only consider one component the Higgs doublet and a lepton. The Higgs doublet however has two components and we can also have processes with a lepton or anti-lepton in the final state.}

In order to obtain the GW spectrum we integrate Eq.~\eqref{eq:rhogw_equation} from $\tM$ until today, i.e., $\tZ$. We can then read off the GW spectrum which is defined as
$h^2\Omgw=h^2\,(1/\rho^0)\,d\rhogwZ/d\ln \kZ$, where $h^2$ is a factor that eliminates the uncertainty in the Hubble expansion rate, $\kZ=2\pi \fgw$ is the GW wavenumber today, $\fgw$ the GW frequency today and $\rhoZ$ the critical energy density today. 
We then get the following expression for the GW spectrum:
\begin{eqnarray}
	h^2\Omgw
	=
	h^2\OmegagammaZ
	\frac{Y^2}{64\pi^2\, }
	\frac{M^2}{ m_{\rm Pl}^2} \, 
	\int_{\tM}^{\tD}dt\,
    \frac{a^4}{ \aZ^4}
    \frac{\rhoN(t)}{\rhogammaZ }
	\, \Theta\left(1-\frac{2k}{M}\right) \,k\, \left(\frac{2k }{M}-2\right)^2 \left(1-\frac{2k }{M}\right),
    \label{eq:OmegaGW_decay}
\end{eqnarray}
where $\Theta$ is the Heaviside step-function, $\rhogammaZ$ is the photon energy density today, $h^2\OmegagammaZ=h^2\rhogammaZ/\rhoZ=2.47\times 10^{-5}$.

In order to derive Eq.~\eqref{eq:OmegaGW_decay} used the fact that the final state particles are mass less.
We have also dropped the terms $\nB(r)$ and $\nF(q)$ that appear in the gain term in Eq.~\eqref{eq:collision_term}. When both momenta $r$ and $q$ are of the order of the mass $M$, i.e., $q,r\simeq M \gg T$, it is clear that $\nB(r),\, \nF(q)\ll1$.
When one of the momenta $r$ or $q$ is much smaller than $M$ it can be comparable to the temperature and the same argument does not hold anymore. However, in that scenario the matrix element suppresses the gain term.
One can see this as follows: $x_{G}=k\frac{M}{2}, x_L=q\frac{M}{2}, x_H=r\frac{M}{2}$ and from energy conservation $2=x_G+x_L+x_H$. Let's say $x_L\ll 1$. Then $2\simeq x_G+x_H$ and $x_G\leq 1$ and $x_H\leq 1$ both have to be close to one. The matrix element squared is $\left|\mathcal{M}\right|^2\sim (1-x_G)(2-x_Gx_L-x_G)\approx (1-x_G)(2-x_G)$ and this is suppressed since $x_G$ is very close to one. 
In the other scenario when $x_H\ll 1$ we have again $2\simeq x_G+x_L$, which implies again that the matrix element squared suppresses the gain term.
When deriving Eq.~\eqref{eq:OmegaGW_decay} we also had to perform the phase-space integral over the matrix element squared. We show the calculation explicitly in appendix~\ref{sec:phase_space_integral}.

In order to get an intuition for how the GW spectrum scales with the parameters $Y$ and $M$ we make an instantaneous decay approximation, i.e. we assume that all right-handed neutrinos decay at $\tD$:
\begin{eqnarray}
	h^2\Omgw
	\simeq
	h^2\OmegagammaZ
	\frac{Y^2}{64\pi^2\, }
	\frac{M^2}{ m_{\rm Pl}^2} \, \frac{\rhoND}{\rhogammaZ }\,
	\frac{\aD^4}{\aZ^4}\frac{M}{2}\,
	\frac{1}{\Gamma}\,
	\frac{2\kD }{M}\left(\frac{2\kD }{M}-2\right)^2 \left(1-\frac{2\kD }{M}\right),
    \label{eq:GW_spectrum_decay_inst}
\end{eqnarray}
with $\kD=\frac{\kZ \aZ}{\aD}$. We can use the last expression to find the peak frequency. The function $f(x)=x(x-2)^2(1-x)$ has a maximum at $x_{\rm max}=0.36$ for $x\leq 1$. The peak frequency is therefore located at
\begin{eqnarray}
	\fgw^{\rm peak}=\frac{x_{\rm max}}{2\pi} \frac{M}{2}\frac{\aD}{\aZ}=7.9\times 10^{12}\,{\rm Hz}\, \left(\frac{M}{10^{15}\,{\rm GeV}}\right)^{1/2}\,\left(\frac{10^{-4}}{Y}\right).
    \label{eq:GW_spectrum_decay_f_peak}
\end{eqnarray}
The magnitude of the GW spectrum evaluated at the peak frequency in the instantaneous decay approximation is
\begin{eqnarray}
	h^2\Omgw(\fgw^{\rm peak})
	=
	2.81\times 10^{-16}\left(\frac{M}{10^{15}\,{\rm GeV}}\right)^2.
    \label{eq:GW_spectrum_decay_amplitude_peak}
\end{eqnarray}

\begin{figure}
	\centering
	\includegraphics[width=0.8\textwidth]{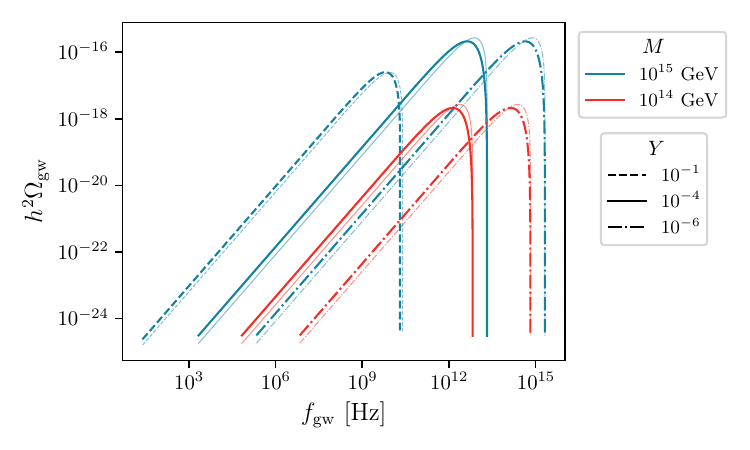}
	\caption{GW spectrum arising from graviton bremsstrahlung during right-handed neutrino decays. We show the GW spectrum for $M=10^{15}$ GeV (blue) and $M=10^{14}$ GeV (red). Different Yukawa couplings are shown as dashed ($Y=10^{-1}$), solid ($Y=10^{-4}$) and dash-dotted ($Y=10^{-6}$). The thick lines show the full solution and the thin lines show the instantaneous decay approximation. In the two scenarios with $Y=10^{-4}$ and $Y=10^{-6}$ there is an era of early matter domination, while for $Y=10^{-1}$ we do not. }
	\label{Fig:GW_spectrum_decay}
\end{figure}

We plot the GW spectrum in Fig.~\ref{Fig:GW_spectrum_decay}. The full solution for the GW spectrum from Eq.~\eqref{eq:OmegaGW_decay} is shown with thick lines and the instantaneous decay approximation, cf. Eq.~\eqref{eq:GW_spectrum_decay_inst} is depicted with thin lines. We observe overall good agreement between the full solution and the instantaneous decay approximation. The blue and red curves depict the cases $M=10^{15}$ GeV and $M=10^{14}$ GeV respectively. We use dashed, solid and dash-dotted curves for the Yukawa couplings $Y=10^{-1}$, $Y=10^{-4}$ and $10^{-6}$. 
For $Y=10^{-4}$ and $10^{-6}$ we are in the limit where we get an era of early matter domination. In those cases the scaling of the peak frequency and magnitude from Eqs.~\eqref{eq:GW_spectrum_decay_f_peak} and~\eqref{eq:GW_spectrum_decay_amplitude_peak} apply.

The dashed curve in Fig.~\ref{Fig:GW_spectrum_decay} shows the GW spectrum for $Y=10^{-1}$ and $M=10^{15}\,$GeV. In that scenario the Yukawa coupling is so large such that we do not get an era of early matter domination anymore. 
As a result the magnitude is reduced compared to the scenario when there is an era of early matter domination with the same $M$ but smaller $Y$. This demonstrates that when calculating the GW spectrum it is imperative to take the correct cosmological evolution into account. 
In the limit of no early matter domination we find the following scaling for the peak frequency $\fgw^{\rm peak}= 8.8\times 10^9\,{\rm Hz}\left(10^{-1}/Y\right)\left(M/(10^{15}\,{\rm GeV})\right)^{1/2}$ and for the magnitude $h^2\Omgw(\fgw^{\rm peak})=2.2\times 10^{-17}\left(M/(10^{15}\,{\rm GeV})\right)^{5/2}\,10^{-1}/Y$. In the scenario $Y=10^{-1}$ and $M=10^{15}$ GeV  scattering processes can wash out the lepton asymmetry. We will study such wash-out scenarios in detail in a future publication but note that even if the right handed neutrinos with $Y=10^{-1}$ and $M=10^{15}$ GeV would not generate a sufficient lepton asymmetry, their decay would still result in the shown GW spectrum.
We do not show the case $Y=10^{-1}$ and $M=10^{14}$ in Fig.~\ref{Fig:GW_spectrum_decay} since it would lead to more wash-out via inverse decays because in such a scenario $\tD<\tM$.

\subsection{Gravitational waves from the thermal plasma}\label{Sec:GW_thermal_plasma}

In this section we calculate the CGMB spectrum in a modified cosmological scenario with an era of early matter domination. Right after the universe reheats, the plasma contains relativistic SM particles as well as right-handed neutrinos. The right-handed neutrinos become non-relativistic at $\tM$ and the universe is radiation dominated until early matter radiation equality at $\tMD$.

The production of GWs starts right after reheating. Scattering processes in the thermal plasma can produce gravitons, which immediately decouple from the thermal plasma. The production is dominated by high temperatures and the dominant processes have two plasma particles in the initial state and one plasma particle and one graviton in the final state, cf. Fig.~\ref{Fig:cosmological_history}. Processes with two gravitons in the final state are sub-dominant, cf. Ref.~\cite{Ghiglieri:2022rfp,Ghiglieri:2024ghm}.

The gain term for graviton production has already been calculated in previous works for all Standard Model processes, as well as for generic beyond-the-Standard-Model (BSM) theories. It can be parameterized as:
\begin{eqnarray}
	\Ggw(k,t)=\frac{16\pi T^4}{m_{\rm Pl}^2 k}\,\hat{\eta}\left(T,\frac{k}{T}\right),
\end{eqnarray}
where the function $\hat{\eta}$ was defined in Ref.~\cite{Ringwald:2020ist}. The diagrams have been primarily calculated in Refs.~\cite{Ghiglieri:2015nfa,Ghiglieri:2020mhm}.

\begin{figure}
	\centering
	\includegraphics[width=0.7\textwidth]{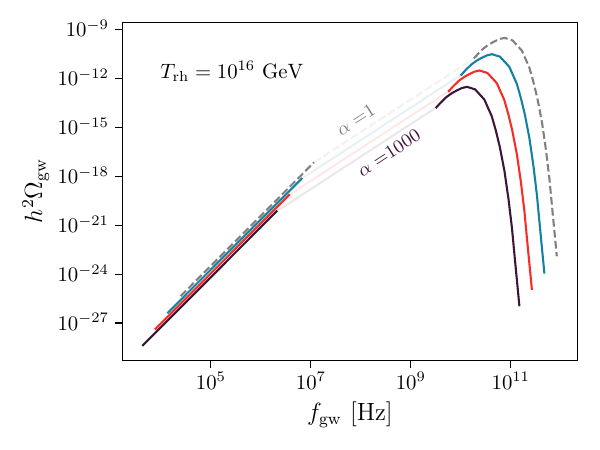}
	\caption{CGMB spectrum with an era of early matter domination. The dashed gray curves depict the scenario with no early matter domination ($\alpha=1$). The larger $\alpha$, the longer the phase of early matter domination. The colored curves show the GW spectrum with an era of early matter domination, blue ($\alpha=10$), red ($\alpha=100$) and black ($\alpha=1000$). The reheating temperature is set to $\Trh=10^{16}$ GeV. }
	\label{fig:CGMB_eMD}
\end{figure}

Integrating Eq.~\eqref{eq:rhogw_equation} gives the GW spectrum
\begin{eqnarray}
	h^2\Omgw=\frac{16 }{\pi}\,h^2\OmegagammaZ\,\frac{1}{\rhogammaZ \aZ^4} \, \int_{\trh}^{\tM} dt \, a^4 \, k^3\,  T^2\left(\frac{T}{m_{\rm Pl}}\right)^2  \hat{\eta}\left(T,\frac{k}{T}\right),
    \label{eq:CGMB}
\end{eqnarray}
where we integrate only until $\tM$ is an excellent approximation because the graviton production is dominated by very high temperatures.
We can perform the integral in Eq.~\eqref{eq:CGMB} analytically:
\begin{eqnarray}
	\Omgw=
	\frac{120 \sqrt{45}}{\pi^{9/2}}\,  \OmegagammaZ\frac{1}{(\gsrh)^{1/2} } \, \frac{\kZ^3}{\TZ^3}\, \left(\frac{\gssZ}{\gsrh}\right)^{1/3}\,\frac{1}{\,\alpha^{1/4}}\, \frac{\Trh}{m_{\rm Pl}}\,\hat{\eta}\left(\Trh,\frac{\kZ}{\TZ}\left(\frac{\gsrh}{\gssZ}\right)^{1/3}\,\alpha^{1/4}  \right) \ , 
    \label{eq:OmGW_CGMB}
\end{eqnarray}
where we used that $\Trh > \TM$ and that between reheating and $\TM$ the $g_*$ functions are temperature independent. Furthermore we have used that the explicit temperature dependence in the $\hat{\eta}$ function is very weak, i.e., $\hat{\eta}(T,\cdot)\simeq \hat{\eta}(\Trh,\cdot)$. 
In addition we introduced the factor $\alpha$ encodes the duration of the early matter domination phase:
\begin{eqnarray}
	\alpha\equiv \frac{\aD}{\aMD}=35.2 \,\left(\frac{M}{10^{15}\,{\rm GeV}}\right)^{2/3}\,\left(\frac{10^{-3}}{Y}\right)^{1/3}
\end{eqnarray}
In order to derive Eq.~\eqref{eq:OmGW_CGMB} we have used Eq.~\eqref{Eq:TD} which yields:
\begin{eqnarray}
	\TM \aM 
	=
	 \TZ \aZ\left(\frac{\aMD}{\aD}\right)^{1/4} \left(\frac{\gssZ}{\gsSM}\right)^{1/3}.
	\label{OmegqGW_step1}
\end{eqnarray}
We plot the GW spectrum in Fig.~\ref{fig:CGMB_eMD} for different values of $\alpha$. The transparent lines connect the particle-like and hydrodynamical regime~\cite{Ghiglieri:2015nfa}.
Note that although the right-handed neutrinos are assumed to be relativistic after reheating, we do not include their contribution to graviton production in the gain term, as it is marginal compared to the SM contribution. The main effect of the right-handed neutrinos on the GW spectrum arises from the phase of early matter domination that they can induce.

The peak frequency of the CGMB spectrum with an era of early matter domination  scales as,
\begin{eqnarray}
    \fgw^{\rm peak}\sim \frac{1}{\alpha^{1/4}}\sim \frac{1}{M^{1/6}}\,Y^{1/12} ,
\end{eqnarray}
and the magnitude of the GW spectrum as:
\begin{eqnarray}
    \Omega_{\rm gw}(\fgw^{\rm peak}) \sim \frac{1}{\alpha}\sim \frac{1}{M^{2/3}}\,Y^{1/3} .
\end{eqnarray}
The peak position shifts to lower frequencies when we make $Y$ smaller.~\footnote{In other modified cosmological scenarios the CGMB can also be shifter to higher frequencies, cf. Ref.~\cite{Muia:2023wru}.} This is in contrast to the scaling of the peak in the graviton bremsstrahlung GW spectrum, where the peak shifts to higher frequencies when $Y$ decreases. It is therefore easier to disentangle the GGMB spectrum and the GW spectrum from bremsstrahlung if $Y$ is smaller.

If there is no phase of early matter domination we have to set $\alpha=1$ in Eq.~\eqref{eq:OmGW_CGMB}. In that scenario we can still distinguish the GW spectrum from a pure SM like spectrum because there will be slightly more effective degrees of freedom $g_*^{\rm rh}=\gsSM+\gsN$. However, this will be a small effect since $\fgw^{\rm peak}\sim (g_*^{\rm rh})^{1/3}$~\cite{Ringwald:2020ist}. 
The by far bigger effect caused from right-handed neutrinos is if they induce a phase of early matter domination which significantly shifts the peak and magnitude of the GW spectrum.

We can absorb the effect of an era of early matter domination, i.e. $\alpha$, in Eq.~\eqref{eq:OmGW_CGMB} in a modified reheating temperature and $\gs$ function: $\tilde{T}_{\rm rh}=\Trh \,\alpha^{3/8}$, $\tilde{g}_{*}^{\rm rh}=\alpha^{3/4}\,\gsrh$. This makes it clear that an era of early matter domination can mimic the effect of more effective degrees of freedom and a larger reheating temperature.
Therefore the CGMB spectrum alone does not include enough information to distinguish between having more relativistic degrees of freedom and a larger reheating temperature or a phase of early matter domination. A possible way around this might be the explicit temperature dependence of the $\hat{\eta}$ function, however such an effect will be small.

\begin{figure}
    \centering
    \includegraphics[width=0.9\linewidth]{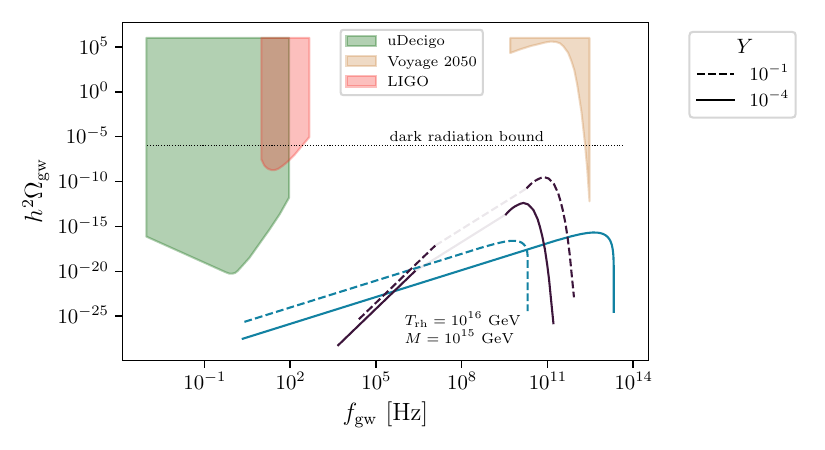}
    \caption{The GW spectrum from graviton bremsstrahlung due to right-handed neutrino decays (blue) and the CGMB spectrum (black). We show the GW spectrum for the parameters $\Trh=10^{16}$ GeV, $M=10^{15}$ GeV. The scenario $Y=10^{-4}$ (solid) leads to an era of early matter domination while $Y=10^{-1}$ (dashed) does not. Smaller $\Trh$ and $M$ would decrease the magnitude of the spectra. We compare the GW spectra with the sensitivities of existing and future GW detection experiments. The LIGO sensitivity is shown in red, while the sensitivity of the proposed uDecigo interferometer is shown in dark green. We also include the projected sensitivity of a futuristic CMB spectral distortion survey from the Voyage 2050 mission (yellow). The dark radiation bound is indicated by the horizontal black dotted line.}
    \label{fig:summary_plot}
\end{figure}

In Fig.~\ref{fig:summary_plot} we compare the CGMB spectrum (black) with the GW spectrum from graviton bremsstrahlung in neutrino decays (blue). 
The low-frequency tail of the CGMB spectrum scales as $\Omega_{\rm gw} \sim \fgw^3$, while the bremsstrahlung spectrum scales linear with the GW frequency $\Omega_{\rm gw} \sim \fgw$. If we could observe both spectra, we can determine the Majorana mass, $M$, and the Yukawa coupling, $Y$, from the bremsstrahlung GW spectrum. 
With that information we could then break the degeneracy among $\gsrh$, $\Trh$, $Y$ and $M$ and infer all four quantities from four observables, namely the two peak frequencies and magnitudes of the GW spectra. In general, observing both spectra will provide valuable insights into particle physics and the early cosmological history of the universe. 

We also confront the calculated GW spectra in Fig.~\ref{fig:summary_plot} with existing and proposed detection schemes. We show the GW spectra for the most optimistic values of reheating temperature and right-handed neutrino mass, i.e., $\Trh=10^{16}\,$ GeV, $M=10^{15}\,$ GeV. Two different Yukawa couplings are considered: $Y=10^{-4}$ (dashed) and $Y=10^{-1}$ (solid).
The sensitivity of the existing LIGO detector~\cite{Abbott:2016xvh} is shown in red, while the sensitivity of the proposed uDecigo detector~\cite{Kuroyanagi:2014qza,Ringwald:2020vei} is shown in dark green.
Ref.~\cite{TitoDAgnolo:2024uku} pointed out that the sensitivity of laser interferometers  can potentially be improved by several orders of magnitude when leveraging the full potential of quantum sensing techniques in the future, which would be needed to reach the low frequency tail of the right-handed neutrino bremsstrahlung GW spectrum.

A stochastic GW background will partly convert into photons in galactic and intergalactic magnetic fields. Future CMB spectral distortion experiments will be sensitive to this photon excess resulting from graviton-photon conversion and can therefore indirectly detect high-frequency GWs. If ESA's large-class science mission Voyage 2050 includes a CMB spectral distortion survey with accuracy on the order of a few nK one can detect GWs~\cite{He:2023xoh} below the dark radiation bound around $10^{12}\,$Hz. 
The dark radiation bound from CMB and BBN measurements is located around $h_0^2\Omgw\lesssim 10^{-6}$~\cite{Ringwald:2020ist}.

The uDecigo sensitivity is roughly 6 orders of magnitude away from the tail of the neutrino decay bremsstrahlung GW spectrum with $M=10^{15}\,$GeV and $Y=10^{-1}$. Voyage 2050 is roughly 4 orders away from the peak of the neutrino decay bremsstrahlung GW spectrum with $M=10^{15}\,$GeV and $Y=10^{-4}$. All in all we can therefore conclude that future improvements of interferometers with for example quantum sensing techniques are needed in order to observe the low frequency tail. Observing the high-frequency peak of the bremsstrahlung GW spectrum with CMB spectral distortion surveys would require a sensitivity improvement beyond Voyage 2050.

Various table-top experiments can also detect high-frequency GWs~\cite{Berlin:2021txa,Domcke:2022rgu,Berlin:2023grv,Kahn:2023mrj,Domcke:2024mfu,Kharzeev:2025lyu}. However, while some of those experiments could potentially probe below the dark radiation bound, they would all require sensitivity improvements in order to detect the GW spectra in Fig.~\ref{fig:summary_plot}. See also Ref.~\cite{Aggarwal:2025noe} for a review.

\section{Conclusion}\label{sec:conclusion}

Leptogenesis is one of the leading paradigms for explaining the baryon asymmetry of the universe. In this scenario, a heavy right-handed neutrino is produced after reheating, and its decay generates a lepton asymmetry, which is subsequently partially converted into a baryon asymmetry.
We have discussed the GW signature of graviton bremsstrahlung from the decay of right-handed neutrinos, whose observation would enable us to see ``leptogenesis in action.'' 
The magnitude of the GW spectrum scales quadratically with the Majorana mass of the decaying right-handed neutrino. The spectrum peaks at high frequencies, with the peak frequency scaling inversely with the Yukawa coupling.
The observation of the bremsstrahlung GW spectrum would provide unambiguous evidence that a heavy particle was produced and subsequently decayed in the early universe. 
While not an ``experimental proof,'' it would constitute strong evidence for leptogenesis, especially when combined with observations of neutrinoless double beta decay and CP violation in neutrino oscillations.

We have also computed the GW spectrum generated from thermal plasma fluctuations in the presence of heavy right-handed neutrinos. The so-called CGMB spectrum can be significantly modified if the right-handed neutrinos induce a phase of early matter domination. Its magnitude and peak frequency are reduced by a factor $\alpha^{1/4}$ and $\alpha$ respectively, where $\alpha$ is the ratio of the scale factor at the end and beginning of early matter domination.
We have shown that an era of early matter domination can mimic additional relativistic degrees of freedom and a higher reheating temperature in the CGMB spectrum. With the help of the graviton bremsstrahlung GW spectrum one could disentangle these two effects in the CGMB.

Note that there is also a potential GW signature from a cosmic string network formed during the breaking of the $U(1)_{B-L}$ gauge symmetry, which generates the mass of the right-handed neutrino.
Other gauge groups such as $SU(2)_R$, Pati--Salam, and flipped $SU(5)$ also give rise to cosmic strings depending on the Higgs particle content. 
Ultimately, the combined study of very low-frequency GWs from cosmic string decays, which can be probed by pulsar timing arrays, and high-frequency GWs in the GHz range and beyond could unveil a detailed cosmological history associated with leptogenesis.

Currently we do not have the technology to observe stochastic GW backgrounds beyond kHz frequencies with sufficient sensitivity. However, future experiments and surveys such as a CMB spectral distortion mission in Voyage 2050 might come quite close. We hope our results will help stimulate further efforts toward developing such capabilities.

\acknowledgments

We thank Jacopo Ghiglieri, Ryan Plestid, Surjeet Rajendran and Tsutomu Yanagida for useful discussions. The work of H.\,M. is supported by the Director, Office of Science, Office of High Energy Physics of the U.S. Department of Energy under the Contract No. DE-AC02-05CH11231, by the NSF grant PHY-2210390, by the JSPS Grant-in-Aid for Scientific Research JP23K03382, MEXT Grant-in-Aid for Transformative Research Areas (A) JP20H05850, JP20A203, Hamamatsu Photonics, K.K, and Tokyo Dome Corportation. In addition, HM is supported by the World Premier International Research Center Initiative (WPI) MEXT, Japan. 
JSE was supported by the National Science Foundation under cooperative agreement 202027 and by the by Japan Science and Technology Agency (JST) as part of Adopting Sustainable Partnerships for Innovative Research Ecosystem (ASPIRE), Grant Number JPMJAP2318.

\appendix

\section{Details of the Matrix element calculation}\label{sec:Matrix_element}

In this appendix we carefully derive the spin and polarization averaged matrix element squared. We specify the Feynman rules in Sec.~\ref{Sec:Feynman_rules}. In Sec.~\ref{Sec:Matrix_element_squared} we then calculate the matrix element squared. We also show that our result is gauge invariant which is a nice consistency check. 
In Sec.~\ref{sec:amplitudes} we repeat the calculation of the matrix element squared with helicity amplitudes and show that we get the same result as in Sec.~\ref{Sec:Matrix_element_squared} where we calculate the matrix element squared by calculating a trace.
In Sec.~\ref{sec:phase_space_integral} we compute the phase space integral over the averaged matrix element squared.
The calculation has already been attempted before in Ref.~\cite{Datta:2024tne} and Ref.~\cite{Choi:2025hqt}.
Our end result for the phase space integral over the averaged matrix element squared does not agree with the results in Refs.~\cite{Datta:2024tne,Choi:2025hqt}. We comment on the disagreement when we discuss our results.

\subsection{Feynman rules}\label{Sec:Feynman_rules}
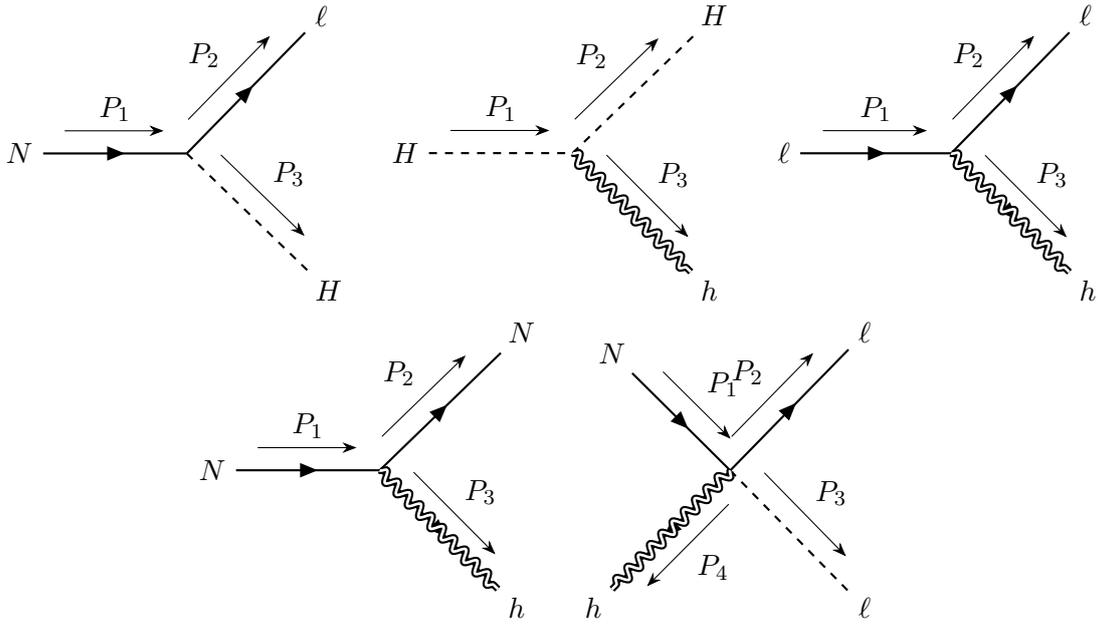
\begin{figure}[t]
	\centering
	\begin{subfigure}[c]{.32\textwidth}
		\begin{tikzpicture}
			\begin{feynman}
				\vertex (a) {\(N\)};
				\vertex [right=2.2cm of a] (b) ;
				\vertex [above right=2.2cm of b] (c){\(\ell \)};
				\vertex [below right=2.2cm of b] (d) {\(H \)};
				\diagram* {
					(a) -- [fermion,thick, momentum=$P_1$] (b), 
					(b) -- [fermion,thick,momentum=$P_2$] (c),
					(b) -- [scalar,thick, momentum=$P_3$] (d)
				};
				~~~~~~\end{feynman}
		\end{tikzpicture}
	\end{subfigure}	
	\begin{subfigure}[c]{.32\textwidth}
		\begin{tikzpicture}
			\begin{feynman}
				\vertex (a) {\(H\)};
				\vertex [right=2.2cm of a] (b) ;
				\vertex [above right=2.2cm of b] (c){\(H \)};
				\vertex [below right=2.2cm of b] (d) {\(h \)};
				\diagram* {
					(a) -- [scalar,thick, momentum=$P_1$] (b), 
					(b) -- [scalar,thick,momentum=$P_2$] (c),
					(b) -- [scalar,thick,momentum=$P_3$] (d),
					(b) -- [thick, double distance=1pt,decoration={snake, amplitude=2pt, segment length=6pt}, decorate] (d)
				};
				~~~~~~\end{feynman}
		\end{tikzpicture}
	\end{subfigure}	
	\begin{subfigure}[c]{.32\textwidth}
		\begin{tikzpicture}
			\begin{feynman}
				\vertex (a) {\(\ell\)};
				\vertex [right=2.2cm of a] (b) ;
				\vertex [above right=2.2cm of b] (c){\(\ell \)};
				\vertex [below right=2.2cm of b] (d) {\(h \)};
				\diagram* {
					(a) -- [fermion,thick, momentum=$P_1$] (b), 
					(b) -- [fermion,thick,momentum=$P_2$] (c),
					(b) -- [fermion,thick,momentum=$P_3$] (d),
					(b) -- [thick, double distance=1pt,decoration={snake, amplitude=2pt, segment length=6pt}, decorate] (d)
				};
				~~~~~~\end{feynman}
		\end{tikzpicture}
		\end{subfigure}\\
		\begin{subfigure}[c]{.32\textwidth}
			\begin{tikzpicture}
				\begin{feynman}
					\vertex (a) {\(N\)};
					\vertex [right=2.2cm of a] (b) ;
					\vertex [above right=2.2cm of b] (c){\(N \)};
					\vertex [below right=2.2cm of b] (d) {\(h \)};
					\diagram* {
						(a) -- [fermion,thick, momentum=$P_1$] (b), 
						(b) -- [fermion,thick,momentum=$P_2$] (c),
						(b) -- [fermion,thick,momentum=$P_3$] (d),
						(b) -- [thick, double distance=1pt,decoration={snake, amplitude=2pt, segment length=6pt}, decorate] (d)
					};
					~~~~~~\end{feynman}
			\end{tikzpicture}
	\end{subfigure}	
	\begin{subfigure}[c]{.32\textwidth}
	\begin{tikzpicture}
		\begin{feynman}
			\vertex (a) {\(N\)};
			\vertex [below right=2.2cm of a] (b) ;
			\vertex [above right=2.2cm of b] (c){\(\ell \)};
			\vertex [below right=2.2cm of b] (d) {\(\ell \)};
			\vertex [below left=2.2cm of b] (e) {\(h \)};
			\diagram* {
				(a) -- [fermion,thick, momentum=$P_1$] (b), 
				(b) -- [fermion,thick,momentum=$P_2$] (c),
				(b) -- [scalar,thick,momentum=$P_3$] (d),
				(b) -- [fermion,thick,momentum=$P_4$] (e),
				(b) -- [thick, double distance=1pt,decoration={snake, amplitude=2pt, segment length=6pt}, decorate] (e)
			};
			~~~~~~\end{feynman}
	\end{tikzpicture}
\end{subfigure}	
	\caption{Feynman rules that we need to calculate all processes that we are interested in. The particle flow of the Higgs ($H$) is always parallel to the momentum. The graviton $h$ is depicted by the doubly wiggly lines and the fermions are depicted by solid lines.}
	\label{fig:Feynman_rules}
\end{figure}

In this section we list all Feynman rules that we need in orde to calculate the relevant diagrams. The diagrams for the vertex Feynman rules are shown in Fig.~\ref{fig:Feynman_rules},
\begin{align}
	N-\ell-H~\text{vertex:}&\qquad -iY P_R\nonumber\\
	H-H-h~\text{vertex:}&\qquad \frac{i\kappa}{2}\left(\eta_{\mu\nu}P_1\cdot P_2-P_{1\mu}P_{2\nu}-P_{1\nu}P_{2\mu}\right)\nonumber\\
	\ell-\ell-h~\text{vertex:}&\qquad  \frac{i\kappa}{8}\left[\eta_{\mu\nu}(\slashed{P}_1+\slashed{P}_2)-(P_1+P_2)_\mu\gamma_\nu-\gamma_\mu(P_1+P_2)_\nu\right]\nonumber\\
	N-N-h~\text{vertex:}&\qquad  \frac{i\kappa}{8}\left[\eta_{\mu\nu}(\slashed{P}_1+\slashed{P}_2-2M)-(P_1+P_2)_\mu\gamma_\nu-\gamma_\mu(P_1+P_2)_\nu\right]\nonumber\\
	N-\ell-H-h~\text{vertex:}&\qquad \frac{i\kappa}{8}\eta_{\mu\nu}(-2 Y)P_R,\nonumber
\end{align}
where we have defined $\kappa=\sqrt{32\pi G}$. We have also used the notation $\slashed{P}=P_\mu\gamma^\mu$, $M$ is the Majorana mass of the right-handed neutrino and $Y$ is the Yukawa coupling. The left ($L$) and right ($R$) handed projectors are $P_{L,R}=(1\mp\gamma^5)/2$.

The fermion-fermion-graviton vertex Feynman rule is consistent with the rule that was given in Ref.~\cite{Holstein:2006bh} (see Eq. (90)). However, it differs by a factor of $2$ compared to~\cite{Choi:1994ax}, which has a factor of $2$ in front of $\eta_{\mu\nu}$. We stick to the result from Ref.~\cite{Holstein:2006bh} which gives us a gauge invariant result in the end. One could use the result from Ref.~\cite{Choi:1994ax}, but then one would have to replace the `$2$' in the $N-\ell-H-h$ vertex with a $4$. In general, we found that one can replace the pre-factor in front of the $\eta_{\mu\nu}$ in the fermion-fermion-graviton vertex by an arbitrary factor if one at the same time re-scales the $N-\ell-H-h$ vertex by the same factor and one obtains the same gauge invariant result.

The propagators are
\begin{align}
	\ell:&\qquad  \frac{i\slashed{P}}{P^2}\nonumber\\
	N:&\qquad  \frac{i(\slashed{P}+M)}{P^2-M^2}\nonumber\\
	H:&\qquad  \frac{i}{P^2}\nonumber
\end{align}

\subsection{Matrix element squared}\label{Sec:Matrix_element_squared}

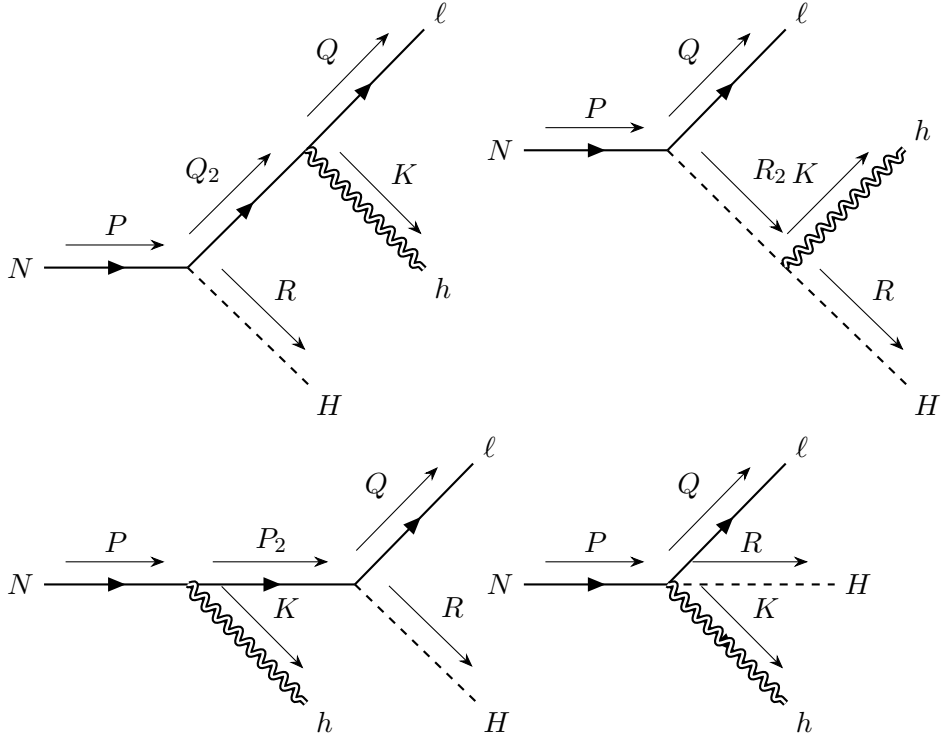
\begin{figure}[t]
	\centering
	\begin{subfigure}[c]{.4\textwidth}
		\begin{tikzpicture}
			\begin{feynman}
				\vertex (a) {\(N\)};
				\vertex [right=2.2cm of a] (b) ;
				\vertex [above right=2.2cm of b] (c);
				\vertex [above right=2.2cm of c] (d) {\(\ell\)};
				\vertex [below right=2.2cm of c] (e) {\(h \)};
				\vertex [below right=2.2cm of b] (f) {\(H \)};
				\diagram* {
					(a) -- [fermion,thick, momentum=$P$] (b), 
					(b) -- [fermion,thick,momentum=$Q_2$] (c),
					(c) -- [scalar,thick,momentum=$K$] (e),
					(c) -- [thick, double distance=1pt,decoration={snake, amplitude=2pt, segment length=6pt}, decorate] (e),
					(c) -- [fermion,thick, momentum=$Q$] (d),
					(b) -- [scalar,thick, momentum=$R$] (f)
				};
				~~~~~~\end{feynman}
		\end{tikzpicture}
	\end{subfigure}	
	\begin{subfigure}[c]{.4\textwidth}
		\begin{tikzpicture}
			\begin{feynman}
				\vertex (a) {\(N\)};
				\vertex [right=2.2cm of a] (b) ;
				\vertex [above right=2.2cm of b] (c) {\(\ell\)};
				\vertex [below right=2.2cm of b] (d) ;
				\vertex [below right=2.2cm of d] (f) {\(H \)};
				\vertex [above right=2.2cm of d] (e) {\(h \)};
				\diagram* {
					(a) -- [fermion,thick, momentum=$P$] (b), 
					(b) -- [fermion,thick,momentum=$Q$] (c),
					(b) -- [scalar,thick,momentum=$R_2$] (d),
					(d) -- [scalar,thick, momentum=$K$] (e),
					(d) -- [thick, double distance=1pt,decoration={snake, amplitude=2pt, segment length=6pt}, decorate] (e),
					(d) -- [scalar,thick, momentum=$R$] (f)
				};
				~~~~~~\end{feynman}
		\end{tikzpicture}
	\end{subfigure}
	\\
	\begin{subfigure}[c]{.4\textwidth}
	\begin{tikzpicture}
		\begin{feynman}
			\vertex (a) {\(N\)};
			\vertex [right=2.2cm of a] (b) ;
			\vertex [right=2.2cm of b] (d) ;
			\vertex [below right=2.2cm of b] (c) {\(h\)};
			\vertex [above right=2.2cm of d] (e) {\(\ell\)};
			\vertex [below right=2.2cm of d] (f) {\(H\)};
			\diagram* {
				(a) -- [fermion,thick, momentum=$P$] (b), 
				(b) -- [fermion,thick,momentum=$P_2$] (d),
				(b) -- [scalar,thick, momentum=$K$] (c),
				(b) -- [thick, double distance=1pt,decoration={snake, amplitude=2pt, segment length=6pt}, decorate] (c),
				(d) -- [fermion,thick, momentum=$Q$] (e),
				(d) -- [scalar,thick, momentum=$R$] (f)				};
			~~~~~~\end{feynman}
	\end{tikzpicture}
\end{subfigure}
	\begin{subfigure}[c]{.4\textwidth}
		\begin{tikzpicture}
			\begin{feynman}
				\vertex (a) {\(N\)};
				\vertex [right=2.2cm of a] (b) ;
				\vertex [above right=2.2cm of b] (c){\(\ell \)};
				\vertex [right=2.2cm of b] (d) {\(H \)};
				\vertex [below right=2.2cm of b] (e) {\(h \)};
				\diagram* {
					(a) -- [fermion,thick, momentum=$P$] (b), 
					(b) -- [fermion,thick,momentum=$Q$] (c),
					(b) -- [scalar,thick,momentum=$R$] (d),
					(b) -- [fermion,thick,momentum=$K$] (e),
					(b) -- [thick, double distance=1pt,decoration={snake, amplitude=2pt, segment length=6pt}, decorate] (e)
				};
				~~~~~~\end{feynman}
		\end{tikzpicture}
	\end{subfigure}	
	\caption{Graviton production processes from right-handed neutrino decay. For the scalar lines particle flux is always chosen in the direction of the momentum arrow.}
	\label{fig:gw_production_from_fermion}
\end{figure}

In this section we compute the amplitude and the matrix element squared that belong to the diagrams that are shown in Fig.~\ref{fig:gw_production_from_fermion}. The two amplitudes corresponding to the diagrams in the top row are called $A_1$ and $A_2$ from left to right. The two diagrams in the bottom row are called $A_3$ and $A_4$ from left to right. Note that we are considering only one component of the lepton doublet (and the corresponding component of the Higgs doublet) and only the lepton final state but not anti-lepton. When considering the total decay rate, the spin averaged squared amplitude must be multiplied by a factor of four.

We refer to the total amplitude as
\begin{align}
	A=A_1+A_2+A_3+A_4,
\end{align}
and we use the notation,
\begin{align}
	A=A_{\mu\nu}\epsilon_\lambda^{\mu\nu*}(K)=\bar{u}^{s'}(Q)\tilde{A}_{\mu\nu}u^s(P)\epsilon_\lambda^{\mu\nu*}(K),
\end{align}
where we call the spin of the incoming right-handed neutrino $s$ and the spin of the outgoing lepton $s'$. The spinors for an incoming particle and an outgoing particle with momentum $P$ and spin $s$ are $\bar{u}^s(P)$ and $u^s(P)$. The polarization tensor for a graviton is $\epsilon_\lambda^{\mu\nu}$ where $\lambda$ is the polarization.

The amplitudes for the four diagrams read
\begin{align}
	\tilde{A}_{1,\mu\nu}
	&=(-iY)\frac{i\kappa}{8}
	P_R\Bigg[
	\left[\eta_{\mu\nu}(\slashed{Q}+\slashed{Q}_2)-(Q+Q_2)_\mu\gamma_\nu-\gamma_\mu(Q+Q_2)_\nu\right]\frac{i\slashed{Q}_2}{Q_2^2}
	\Bigg],
	\\
	\tilde{A}_{2,\mu\nu}&=
	(-iY)\frac{i\kappa}{8}
	P_R\Bigg[
	4 \frac{i}{R_2^2}\left(\eta_{\mu\nu}R\cdot R_2-R_{\mu}R_{2\nu}-R_{\nu}R_{2\mu}\right)
	\Bigg],
	\\
	\tilde{A}_{3,\mu\nu}&=
	(-iY)\frac{i\kappa}{8}
	P_R\Bigg[
	\frac{i(\slashed{P}_2+M)}{P_2^2-M^2}\left[\eta_{\mu\nu}(\slashed{P}+\slashed{P}_2-2M)-(P+P_2)_\mu\gamma_\nu-\gamma_\mu(P+P_2)_\nu\right]
	\Bigg],
	\\
	\tilde{A}_{4,\mu\nu}&=
	(-iY)\frac{i\kappa}{8}
	P_R\Bigg[
	\eta_{\mu\nu}(-2 i)
	\Bigg].
\end{align}
The matrix element squared is 
\begin{align}
	\left|\mathcal{M}\right|^2=|A|^2=(\bar{u}^{s'}(Q)\tilde{A}_{\mu\nu}u^s(P))
	(\bar{u}^{s}(P)\gamma^0\tilde{A}^\dagger_{\alpha\beta}\gamma^0u^{s'}(Q))\epsilon^{\lambda\alpha\beta}(K)\epsilon^{\lambda\mu\nu*}(K).
\end{align}
We want to average over the graviton polarization $\lambda$ as well as over the spin of the initial and final states.
Averaging over the graviton polarization yields
\begin{align}
	\sum_{\lambda}\left|\mathcal{M}\right|^2=(\bar{u}^{s'}(Q)\tilde{A}_{\mu\nu}u^s(P))
	(\bar{u}^{s}(P)\gamma^0\tilde{A}^\dagger_{\alpha\beta}\gamma^0u^{s'}(Q))L^{\alpha\beta\mu\nu}(K)
\end{align}
with
\begin{align}
	L^{\alpha\beta\mu\nu}=\frac{1}{2}\left(P^{\alpha\mu}P^{\beta\nu}+P^{\alpha\nu}P^{\beta\mu}-P^{\alpha\beta}P^{\mu\nu}\right)
\end{align}
and
\begin{align}
	P^{\mu\nu}=\eta^{\mu\nu}-\frac{K^\mu n^\nu+K^\nu n^\mu}{K\cdot n}-\frac{K^\mu K^\nu}{(K\cdot n)^2},
	\label{eq:Pmunu}
\end{align}
where $n$ is a gauge dependent vector that should drop out in the end. 

Next we also average over $s$ and $s'$,
\begin{align}
	\left|\mathcal{M}\right|^2_{\rm av}\equiv\frac{1}{2}\sum_{\lambda s s'}\left|\mathcal{M}\right|^2=\frac{1}{2}L^{\alpha\beta\mu\nu}(K)\,\,
	{\rm Tr}\left[\slashed{Q}\,\tilde{A}_{\mu\nu}\,
	(\slashed{P}+M)\,(\gamma^0\tilde{A}^\dagger_{\alpha\beta}\gamma^0)
	\right],
\end{align}
where we have used the relations,
\begin{align}
	\sum_su_b^s(P)\bar{u}^s_c(P)=(\slashed{P}+M)_{bc},
	\label{eq:ubaru_1}
\end{align} 
and 
\begin{align}
	\sum_su_d^s(Q)\bar{u}^s_a(Q)=(\slashed{Q})_{da}.
	\label{eq:ubaru_2}
\end{align}

While it is possible to give a co-variant form of the averaged matrix element squared, we do not provide such an equation here. Instead we directly evaluate the expression in the rest frame of the massive right-handed neutrino that decays. The momentum four vectors of the right-handed neutrino, the graviton and the lepton are respectively, 
\begin{align}
	P=M\begin{pmatrix}
		1\\0\\0\\0
	\end{pmatrix},
    ~~
    K=x_G \frac{M}{2}\begin{pmatrix}
		1\\0\\0\\1
	\end{pmatrix},
    Q=x_L \frac{M}{2}\begin{pmatrix}
		1\\\sin\theta\\0\\\cos\theta
	\end{pmatrix}.
	\label{eq:momenta}
\end{align}
Note that the choice that $K$ propagates along the $z$-direction is completely arbitrary. Similarly the choice that $q$ is in the $xz$-plane is arbitrary. Both choices can be made without loss of generality. 
Momentum conservation ($P=K+Q+R$) fixes the momentum of the Higgs
\begin{align}
	R=\frac{M}{2}\begin{pmatrix}
		2-x_G-x_L\\ -x_L\sin\theta\\0\\-x_G-x_L\cos\theta
	\end{pmatrix}=
	\frac{M}{2}\begin{pmatrix}
		x_H\\ -x_L\sin\theta\\0\\-x_G-x_L\cos\theta
	\end{pmatrix},
	\label{eq:r}
\end{align}
where we have defined $x_H=2-x_G-x_L$.

We can relate $\theta$ to $x_L$ and $x_G$ and reduce the number of free parameters to two
\begin{align}
	\cos\theta=\frac{2-2x_G-2x_L+x_G x_L}{x_G x_L}.
    \label{eq:theta}
\end{align}
When we plug this parametrization into the equation for the spin and polarization averaged matrix element squared we obtain
\begin{align}
	\left|\mathcal{M}\right|^2_{\rm av}=\frac{1}{2}\left|Y\right|^2\left(\frac{\kappa}{8}\right)^2\,\frac{16 M^2 (1-x_G) (2-x_G-x_G x_L)}{x_G^2}.
	\label{eq:Msq_averaged}
\end{align}
The expression in Eq.~\eqref{eq:Msq_averaged} is manifestly gauge invariant, i.e., it does not depend on the vector $n$ that appears in the projectors in Eq.~\eqref{eq:Pmunu}.

\subsection{Helicity Amplitudes}\label{sec:amplitudes}

It is instructive to study amplitudes for each helicity combination so that we understand the process with confidence. The heavy right-handed neutrino is at rest and can have its spin along the $z$-axis either $S_z = +\frac{1}{2}$ or $S_z = -\frac{1}{2}$. Note that the helicity of the leptons (anti-leptons) is always $-\frac{1}{2}$ ($+\frac{1}{2}$). The remaining choice is the helicity of the graviton $h=\pm 2$. Therefore, there are only four combinations of helicities. Remember in Eq.~\eqref{eq:momenta}, we chose the coordinate system where the graviton goes in the positive $z$ direction. We use the same $z$ axis as the quantization axis of the spin of the initial state right-handed neutrino. 

The decay amplitudes without the graviton emission for the positive $S_z = \frac{1}{2}$ and negative spin $S_z = -\frac{1}{2}$ of the right-handed neutrino is 
\begin{align}
    A_p &= i Y M x_L^{1/2} \sin \frac{\theta}{2}, \nonumber \\
    A_m &= -i Y M x_L^{1/2} \cos \frac{\theta}{2}.
    \label{eq:A}
\end{align}
The polar angle $\theta$ is that of the lepton and its dependence is easy to understand with the angular momentum conservation. When $\theta=0$, the lepton goes in the positive $z$-direction with negative helicity, contributing $S_z = - \frac{1}{2}$. It does not match a positive spin of the initial right-handed neutrino and hence $A_p$ vanishes, while does match a negative spin and hence $A_m$ is maximized. The case $\theta=\pi$ can also be understood in a similar way. 

For the graviton emission amplitudes, we use the polarization tensor of the graviton
\begin{align}
    \epsilon^{\mu\nu}_\pm 
    &= \frac{1}{2} \left( \begin{array}{cccc}
        0 & 0 & 0 & 0\\
        0 & 1 & \pm i & 0\\
        0 & \pm i & -1 & 0\\
        0 & 0 & 0 & 0
    \end{array}
    \right) .
\end{align}
Remember that for the emission diagram, we need to use $\epsilon^{\mu\nu *}_\pm$.

For a fixed final state with a lepton (charged or neutral) and a corresponding Higgs within the doublets, the amplitudes are
\begin{align}
	A_{pp} &    = i \frac{\kappa}{2} \frac{1-x_G}{x_G} Y M x_L^{1/2} \sin \frac{\theta}{2} 
    = i Y \frac{\kappa}{2} M x_G^{-3/2} (1-x_G)(1-x_H)^{1/2}, \\
	A_{pm} &
    = - i \frac{\kappa}{2} \frac{1-x_G}{x_G} Y M x_L^{1/2} \cos \frac{\theta}{2}
    = - i Y \frac{\kappa}{2} M  x_G^{-3/2} (1-x_G)^{3/2} (1-x_L)^{1/2}, \\
	A_{mp} &
    = i \frac{\kappa}{2} \frac{1-x_G}{x_G} Y M x_L^{1/2} \sin \frac{\theta}{2} 
    = i Y \frac{\kappa}{2} M  x_G^{-3/2} (1-x_G) (1-x_H)^{1/2}, \\
	A_{mp} &
    = - i \frac{\kappa}{2} \frac{1}{x_G} Y M x_L^{1/2} \cos \frac{\theta}{2}
    = - i Y \frac{\kappa}{2} M  x_G^{-3/2} (1-x_G) ^{1/2}(1-x_L)^{1/2},
\end{align}
where the first subscript refers to graviton helicity $p$ for $+2$ and $m$ for $-2$, while the second one to right-handed neutrino spin $p$ for $+1/2$ and $m$ for $-1/2$. We have explicitly checked the gauge invariance of the amplitudes by changing $\epsilon^{\mu\nu} \rightarrow \epsilon^{\mu\nu} + K^\mu \xi^\nu + K^\nu \xi^\mu$ with an arbitrary vector $\xi^\mu$. 

The dependence on the graviton energy fraction $x_G$ makes sense in the following way. When $x_G \rightarrow 1$, Eq.~\eqref{eq:theta} says $\cos\theta \rightarrow -1$. Namely it is a kinematic configuration where the graviton goes to the positive $z$ axis with the maximum momentum, while the lepton and Higgs go in the negative $z$ direction sharing the opposite momentum. Namely this limit is where all three particles are collinear. In this limit, the final state has the total angular momentum $S_z = h_G + \frac{1}{2} $ along the positive $z$ axis compared to $S_z = \pm \frac{1}{2}$ of the initial state right-handed neutrino. How much this configuration would violate the angular momentum conservation is shown in the Table~\ref{tab:DeltaSz}. It is clear that the decay amplitudes above have the behavior $(1-x_G)^{|\Delta J_z|/2}$ so that they vanish in the limit $x_G \rightarrow 1$ because of the conservation law of angular momentum. 

\begin{table}[htb]
	\begin{center}
	\begin{tabular}{|c|c|c|c|}
		\hline
		$h_G$ & $(J_z)_i = (s^N_z)$ & $(J_z)_f = h_G + \frac{1}{2}$ & $\Delta J_z = (J_z)_f - (J_z)_i $   \\ \hline
		$+2$ & $+\frac{1}{2}$ & $+\frac{5}{2}$ & $+2$  \\ \hline
		$+2$ & $-\frac{1}{2}$ & $+\frac{5}{2}$ & $+3$  \\ \hline
		$-2$ & $+\frac{1}{2}$ & $-\frac{3}{2}$ & $-2$ \\ \hline
		$-2$ & $-\frac{1}{2}$ & $-\frac{3}{2}$ & $-1$ \\ \hline
	\end{tabular}
    \raisebox{-42pt}{
    	\includegraphics[width=0.2\textwidth]{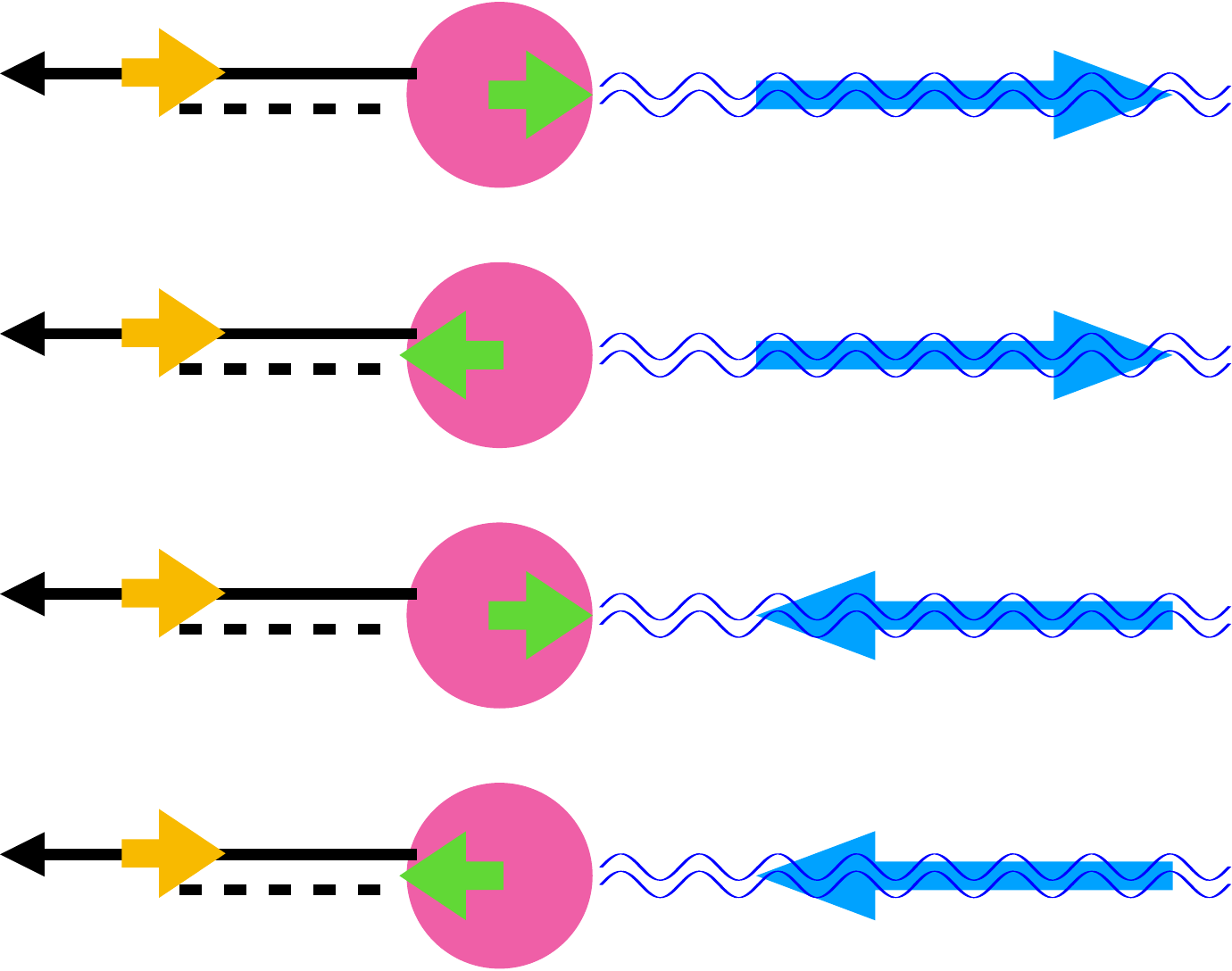}
        }
	\end{center}
	\caption{Change in the angular momentum in the $+z$ direction when $x_G=1$ and hence the graviton goes in the $+z$ direction while the lepton with negative helicity goes into the $-z$ direction. The angular momentum along the $z$ direction in the initial state is the spin of the right-handed neutrino $(J_z)_i = (s^N_z)$.  On the other hand, the angular momentum of the final state along the $z$ direction is $(J_z)_f = h_G + \frac{1}{2}$ because of the negative helicity lepton moving in the negative $z$ direction. The last column shows the change in the angular momentum, $h_G + \frac{1}{2} - s_N$ where $s_N$ is the spin of the right-handed neutrino in the $+z$ direction. }
	\label{tab:DeltaSz}
\end{table}

It is easy to see that the spin-averaged squared amplitude Eq.~\eqref{eq:Msq_averaged} is reproduced from these helicity amplitudes. It vanishes as $(1-x_G)$ in the limit $x_G \rightarrow 1$ because every helicity amplitude vanishes in this limit due to the violation of angular momentum conservation where the least suppression is due to that with $\Delta J_z = 1$. 

A similar consideration should be possible for $x_L \rightarrow 1$ or $x_H \rightarrow 1$ when choosing the quantization axis for the right-handed neutrino spin along the lepton or Higgs momentum. We do not attempt it here.

We can also see that our amplitude agrees with Weinberg's soft graviton theorem \cite{Weinberg:1965nx} valid for $x_G \rightarrow 0$. The theorem says the soft graviton amplitude is proportional to that without a graviton emission up to a factor
\begin{align}
    \eta \frac{\kappa}{2} 
    \frac{X^\mu X^\nu \epsilon^*_{\mu\nu}}{X \cdot K},
\end{align}
where $X^\mu$ is the four momentum of the particle that emits the soft graviton, $\epsilon_{\mu\nu}$ is the graviton polarization tensor, and $\eta=+1$ for outgoing and $-1$ for incoming particles. This factor vanishes for the right-handed neutrino, while it is
\begin{align}
    \eta \frac{\kappa}{2} 
    \frac{Q^\mu Q^\nu \epsilon^*_{\mu\nu}}{Q \cdot K}
    =\frac{\kappa}{2} \frac{(1-x_G)(1-x_L)}{x_G^2}
\end{align}
for the lepton, and
\begin{align}
    \eta \frac{\kappa}{2} 
    \frac{R^\mu R^\nu \epsilon^*_{\mu\nu}}{R \cdot K}
    =\frac{\kappa}{2} \frac{(1-x_G)(1-x_H)}{x_G^2}
\end{align}
for the Higgs. Summing them, we have the overall factor of
\begin{align}
    \frac{\kappa}{2} \frac{(1-x_G)}{x_G}\ .
\end{align}
Comparing to the amplitudes without a graviton emission \eqref{eq:A},
the soft graviton theorem is satisfied in the limit $x_G \rightarrow 0$ (in fact, exact for $A_{pp}, A_{pm}, A_{mp}$). 

It is important that we understand the $1/x_G$ singularity in the amplitude as it appears due to the soft graviton. Given that the right-handed neutrino decays in a hot thermal bath, one may be concerned that the singularity may be softened and the tail of the graviton energy density at lower frequencies shown in Fig.~\ref{fig:summary_plot} may be cutoff. In fact, if the singularity were due to a collinear singularity, a thermal mass for the lepton and Higgs would remove such a singularity. On the other hand, the soft singularity persists even in the presence of thermal effects because the graviton is so weakly coupled and remains massless in the thermal bath. The absence of a collinear singularity is also obvious from the fact that the decay rate has only a single logarithmic divergence, not a double log. 

In fact, we do not expect a collinear singularities in the graviton emission. Collinear singularities arise when a massless particle emits another massless particle, 
\begin{align}
	X^\mu &= E(1,0,0,1) = Y^\mu + Z^\mu, \\
    Y^\mu &= xE(1,\sin\theta,0,\cos\theta), \\
    Z^\mu &= E(1-x,-x\sin\theta,0,-x\cos\theta),
\end{align}
where the $Y$ is the four-momentum of the emitted particle such as a photon or a graviton, while the off-shell particle of momentum $Z$ is actually on-shell in the collinear limit,
\begin{align}
	Z^2 = (X-Y)^2 = 2X\cdot Y = x E^2 (1-\cos\theta) \rightarrow 0 \quad (\theta \rightarrow 0).
\end{align}
Because the propagator for this off-shell particle is $1/Z^2 \propto 1/(1-\cos\theta) \sim 2/\theta^2$, it leads to a singularity, while a finite mass of the particle removes the singularity as $1/(1-\beta\cos\theta)$. For a gauge boson emission like photon or gluon from a scalar particle as an example, there is a numerator that softens the singularity because the angular momentum conservation is violated by one in the collinear limit. The polarization vector is
\begin{align}
	\epsilon^\mu (Y) = \frac{1}{\sqrt{2}} (0, \pm \cos\theta, i, \mp \sin\theta)
\end{align}
and 
\begin{align}
	e (X+Z)^\mu \epsilon_\mu^*
	= \pm \frac{1}{\sqrt{2}} E \sin\theta
\end{align}
which vanishes in the collinear limit. As a result, the singularity is $\sin\theta/(1-\cos\theta) \sim 2/\theta$, not $2/\theta^2$. On the other hand for the graviton emission, the corresponding polarization tensor is
\begin{align}
	\epsilon^{\mu\nu}_\pm(Y) = \frac{1}{2}\left( \begin{array}{cccc}
	0 & 0 & 0 & 0\\
	0 & \cos^2 \theta & \pm i \cos\theta & -\cos\theta\sin\theta \\
	0 & \pm i \cos\theta & -1 & \mp i \sin\theta \\
	0 & -\cos\theta\sin\theta & \mp i \sin\theta & \sin^2 \theta
	\end{array} \right),
\end{align}
and the numerator is
\begin{align}
	\kappa Z_\mu \epsilon_\pm^{\mu\nu} X_\nu = \frac{1}{2} \kappa E^2 \sin^2 \theta
\end{align}
because the angular momentum conservation is violated by two units in the collinear limit, where the amplitude $\sin^2 \theta / (1-\cos\theta) \rightarrow 2$ is regular.

For the low-frequency tail of the CGMB spectrum from the thermal plasma~\cite{Ghiglieri:2015nfa}, a hydrodynamical calculation was required. For the case of the neutrino decay signal, this is not needed. Firstly, the dominant contribution to the spectrum comes from the out-of-equilibrium decay of non-relativistic neutrinos decoupled from the SM plasma. The gravitons are sourced from the local decays rather than large-scale density fluctuations, and once emitted are free-streaming. Hence our calculation should be accurate for both the high- and low-frequency regimes.

Our result does not agree with \cite{Datta:2024tne} where the spin-averaged squared amplitude vanishes as $(1-x_G)^2$ in the $x_G \rightarrow 1$ limit. (Note that their $x$ is our $x_G/2$.) It is in contradiction to the presence of $\Delta J_z = 1$ amplitude which gives rise to the $(1-x_G)$ behavior. Given that we verified our amplitudes both in the $x_G \rightarrow 1$ limit by angular momentum violation and $x_G \rightarrow 0$ by the soft graviton theorem, we are quite confident about our result. The authors in Ref.~\cite{Choi:2025hqt} also discuss the graviton emission from right-handed neutrino decays. They, however, seem to have copied the expression from a scalar particle decay \cite{Hu:2024awd} which is not appropriate for right-handed neutrinos.

\subsection{Phase space integral}\label{sec:phase_space_integral}
We evaluate the three body phase space integral
\begin{align}
	&~\int d\Phi_3(K,Q,R)f(P,K,Q,R)\nonumber\\
    &=\int \frac{d^3k}{(2\pi)^3\, 2k}\, \int \frac{d^3q}{(2\pi)^3\, 2q}\, \int \frac{d^3r}{(2\pi)^3\, 2r}\, (2\pi)^4\,\delta(P-K-Q-R) f(P,K,Q,R),
\end{align}
where we keep the function $f$ under the integral general for a moment.
We work in the rest frame of the decaying particle, i.e., $P=(M,0,0,0)^T$.
We use the delta function to evaluate the $d^3r$ integral. We then set $\bm{r}=-\bm{k}-\bm{q}$. By re-writing the remaining delta function we can also evaluate the $dq$ integral
\begin{align}
	&~\int d\Phi_3(K,Q,R)f(P,K,Q,R)\nonumber\\
	&=\int \frac{dk\, k d\cos\theta_k d\phi_k}{(2\pi)^3\, 2}\, \int \frac{ x_0 d\cos\theta_q d\phi_q}{(2\pi)^2\, 2}\,  \frac{1}{\, 2r}\,\frac{1}{|g'(x_0)|} f(P,K,Q,R)
\end{align}
with
\begin{eqnarray}
	g(x)=M-k-q-\sqrt{k^2+q^2+2k q\cos\theta_{kq}}
\end{eqnarray}
and the zero of $g(x)$ is
\begin{eqnarray}
	x_0=\frac{M}{2}\frac{ M-2 k}{ M+k \cos (\theta_{kq})-k},
\end{eqnarray}
where $\theta_{kq}$ is the angle between $\bm{k}$ and $\bm{q}$. 
Note that $x_0\sim (M-2k)$. Since $x_0>0$ because it is a magnitude of a vector, we need $k\leq\frac{M}{2}$. 
The four vectors are given by
\begin{eqnarray}
	K=\begin{pmatrix}
		k\\
		k \hat{\bm{e}}_k
	\end{pmatrix}, \quad
	Q=\begin{pmatrix}
		x_0\\
		x_0 \hat{\bm{e}}_q
	\end{pmatrix}, \quad
	R=P-K-Q=
	\begin{pmatrix}
		M-k-x_0\\
		-k \hat{\bm{e}}_k-x_0 \hat{\bm{e}}_q
	\end{pmatrix}.	
\end{eqnarray}
Since we are in the rest frame of the decaying particle and we expect that no matter in what direction $\hat{\bm{e}}_k$ points the result for the $q$ integral will always be the same. Intuitively, it is also clear that the result should always be the same for every $\phi_q$. Therefore we can just set
\begin{eqnarray}
	K=\begin{pmatrix}
		k\\
		0\\0\\k
	\end{pmatrix}, \qquad
    Q=\begin{pmatrix}
		x_0\\
		x_0\sin\theta_q\\0\\x_0\cos\theta_q
	\end{pmatrix},
\end{eqnarray}
where we have used that $\hat{\bm{e}}_k\cdot \hat{\bm{e}}_q=\cos\theta_{kq}=\cos\theta_q$.
We can now perform three angle integrals in the phase space integral
\begin{align}
	\int d\Phi_3(K,Q,R)f(P,K,Q,R)
	&=
	\frac{8\pi^2}{(2\pi)^5\, 8}\int_{0}^{M/2} dk\,   \int_{1}^{-1} d\cos\theta_q \,\,  \frac{k x_0}{r}\,\frac{1}{|g'(x_0)|} f(P,K,Q,R).
	\label{eq:final_result_no_angle_simplifications}
\end{align}
Next we transformation to the integration variables $x_G$ and $x_L$,
\begin{eqnarray}
	x_G=k\frac{2}{M}, \qquad
	x_L=x_0\frac{2}{M}.
\end{eqnarray}
Plugging everything into the phase space integral we get
\begin{align}
	\int d\Phi_3(K,Q,R)f(P,K,Q,R)
	=
	\frac{M^2}{128\pi^3\, }\int_0^{1} \,dx_G   \int_{1-x_G}^{1}dx_L\,  f(P,K,Q,R),
	\label{eq:final_result_angle_approximation}
\end{align}
where we have used that $x_G+x_L<2$ which is evident from the integration boundaries.

We now use Eq.~\eqref{eq:final_result_angle_approximation} in order to evaluate the phase space integral with $f=|\mathcal{M}|^2_{\rm av} k$,
\begin{align}
	&~\int d\Phi_3(K,Q,R)\left|\mathcal{M}\right|^2_{\rm av} k\nonumber\\
	&=
	\frac{1}{2}\frac{M^2}{128\pi^3\, }\left|Y\right|^2\left(\frac{\kappa}{8}\right)^2\,\int_0^{1} \,dx_G   \int_{1-x_G}^{1}dx_L\,  \frac{16 M^2 (x_G-1) (x_G x_L+x_G-2)}{x_G^2}x_G\frac{M}{2}\nonumber\\
	&=
	\frac{1}{2}\frac{M^5}{32\pi^3\, }\left|Y\right|^2\left(\frac{\kappa}{8}\right)^2\,\int_0^{1} \,dx_G \,   (x_G-2)^2 (1-x_G).
\end{align}

\bibliographystyle{JHEP}
\bibliography{bibliography}

\end{document}